\documentclass[reprint,
 amsmath,amssymb,
aip, jcp
]{revtex4-2}

\pdfoutput=1
\usepackage{makeidx}
\makeindex
\usepackage{graphicx}
\usepackage{dcolumn}
\usepackage{bm}
\usepackage{hyperref}

\usepackage[
text={7in,10in},centering,
]{geometry}

\usepackage{amssymb}
\usepackage[caption=false]{subfig}
\usepackage{booktabs}

\newcommand{\myciteauthor}[1]{{\protect\NoHyper\citeauthor{#1}\endNoHyper}}
\DeclareGraphicsExtensions{.pdf,.png,.jpg}

\begin{document}

\widetext{\noindent This article may be downloaded for personal use only. Any other use requires prior permission of the author and AIP Publishing. This article appeared in J. Chem. Phys. 152, 134103 (2020) and may be found at \url{https://doi.org/10.1063/5.0002246}.}

\title{Plane wave basis set correction methods for RPA correlation energies }

\author{Stefan Riemelmoser}
\email[Author to whom correspondence should be addressed: \\]{stefan.riemelmoser@univie.ac.at} 
 \affiliation{%
 Faculty of Physics and Center for Computational Materials Science, University of Vienna, Sensengasse 8/12,
A-1090 Vienna, Austria 
}
\author{Merzuk Kaltak} 
\affiliation{%
VASP Software GmbH, Sensengasse 8/17,
A-1090 Vienna, Austria 
}
\author{Georg Kresse}%
\affiliation{%
Faculty of Physics and Center for Computational Materials Science, University of Vienna, Sensengasse 8/12,
A-1090 Vienna, Austria 
}%




\date{\today }\vspace{\baselineskip}

\begin{abstract}
Electronic correlation energies from the random-phase approximation converge slowly with respect to the plane wave basis set size. We study the conditions under which a short-range local density functional can be used to account for the basis set incompleteness error. Furthermore, we propose a one-shot extrapolation scheme based on the Lindhard response function of the homogeneous electron gas. The different basis set correction methods are used to calculate equilibrium lattice constants for prototypical solids of different bonding types. 

\end{abstract}
%
\maketitle
%
%
\section{\label{sec:Riemelmoser2020_1} Introduction}
The random-phase approximation (RPA) for the electronic correlation energy was developed during the beginnings of many-body perturbation theory in the 1950s. The detailed analytical studies of the homogeneous electron gas (HEG) in the RPA have contributed greatly to the understanding of many-body effects in solid state physics such as plasmons, screening and electronic correlation in general.\cite{Pines1989} However, large scale practical applications of the RPA have been possible only in the last one or two decades due to increasing computer power.  

In the 1970s, the RPA was formulated in the framework of density functional theory (DFT)\cite{Hohenberg1964,Kohn1965} by Langreth and Perdew via the adiabatic connection formalism. \cite{Langreth1975,Langreth1977} It can be categorized as a fifth rung functional (i.e. includes unoccupied orbitals) on Jacob's ladder. \cite{Perdew2001} In the early 2000s, renewed interest was drawn to the RPA following the work of \myciteauthor{Yan2000}\cite{Yan2000,*Yan2010} and \myciteauthor{Furche2001}.\cite{Furche2001} Most importantly, the RPA  - unlike lower rung functionals - describes long-range effects like van der Waals - bonding and dispersion very well. Furthermore, the RPA is compatible with exact exchange and behaves well in the metallic limit, where other methods such as M{\o}ller-Plesset perturbation theory diverge.\cite{Harl2008,Harl2010} Finally, the adiabatic connection formalism provides a natural way to construct beyond-RPA theories, see e.g. Ref. \onlinecite{Furche2005}. The recent development has been summarized in the review of \myciteauthor{Ren2012},\cite{Ren2012} which also includes derivations of the standard RPA expressions.

One limiting factor of the RPA, like for other methods based on many-body perturbation theory, is the slow convergence of the correlation energy with respect to the basis set size. In a canonical plane wave implementation, the RPA scales at least as $N_{\rm PW}^3$ with respect to the number of plane waves $N_{\rm PW}$.\cite{Kaltak2014,Kaltak2014a,Rojas1995} On the other hand, the basis set incompleteness error decays only as $1/N_{\rm PW}$. \cite{Harl2008,Klimes2014a,Gulans2014} The reason for this slow convergence is connected to the electronic cusp condition,\cite{Kato1957,Helgaker2014} which states that the exact many-body wave function has a kink at electron coalescence. This (integrable) UV - divergence is caused by the $1/r$ - singularity of the Coulomb potential, or equivalently by its  $1/q^2$ - high momentum behavior. The cusp-related convergence problem is not limited to plane waves, but applies also to local basis sets. \cite{Furche2001}

The most straightforward strategy to reduce the computational cost is to perform a basis set extrapolation. \cite{Harl2008,Gulans2014} Alternatively, range-separated DFT \cite{Toulouse2004} circumvents the cusp condition directly by replacing the short-range part of the Coulomb interaction, which includes the singularity, by DFT. 
Range-separation has been applied to the RPA by Toulouse \textit{et al.},\cite{Toulouse2009,Toulouse2010} Janesko \textit{et al.}\cite{Janesko2009,Janesko2009a,*Janesko2013,Janesko2009b,*Janesko2010} and \myciteauthor{Bruneval2012},\cite{Bruneval2012} who have all reported a significant reduction of computational cost. 

Range-separation has possibly further advantages. The RPA has well known shortcomings in describing short- and mid-range correlation effects, such as the prediction of an unphysical bump in the dissociation curve of $\text{Be}_2$.\cite{Toulouse2009} Unsurprisingly, it is thus possible to improve on full-range RPA by creating range-separated hybrid functionals. The price to pay is the introduction of empirical parameters into the theory. Inverse range-separation, which aims to cure IR divergences, has been applied extensively to the exchange energy, e.g. in the popular HSE hybrid functional. \cite{Heyd2003,*Heyd2006}

Range-separation schemes up to now have been based mostly on the error function. In this paper, we drop this constraint and investigate the performance of alternative long-range potentials. First, we recall the wave vector decomposition method of old\cite{Nozieres1958} to study range-separated RPA based on momentum cutoffs. Second, we propose an optimized long-range potential (``squeezed Coulomb kernel'') that performs an implicit basis set extrapolation.

In Sec. \ref{sec:Riemelmoser2020_2}, we discuss our range-separation scheme, and study exacts limits for range-separation based on wave vector decomposition. Then, we use the results for the low-density HEG to construct the squeezed Coulomb kernel. In Sec. \ref{sec:Riemelmoser2020_3}, we present numerical studies for the HEG and discuss the analytical representations of our local density functionals. In Sec. \ref{sec:Riemelmoser2020_4}, we show test calculations for a small set of prototypical materials and compare the different basis set correction methods.  Conclusions are drawn in Sec. \ref{sec:Riemelmoser2020_5}. The discussion is restricted to the case of non-spin-polarized electrons. Unless stated otherwise, Hartree units are used throughout the work. 

\section{\label{sec:Riemelmoser2020_2} Theory}
\subsection{Range-separated density functional theory}

Since we are interested in a simple basis set correction method for post-DFT RPA calculations, we adopt the range-separation scheme of Bruneval. \cite{Bruneval2012} Generally, the Coulomb kernel is decomposed in a long-range and a complementary short-range part 
\begin{equation}
V = V^{\rm LR} (\mu) + V^{\rm SR} (\mu), 
\end{equation}
where $\mu$ is a tunable range-separation parameter and cuts of the Coulomb potential at a cutoff radius $r_{\rm cut} \approx 1/\mu$. The long-range version of the RPA correlation energy $E_{\rm c}^{\rm RPA, LR}$ is obtained by replacing $V$ by $V^{\rm LR}$ in the text-book expression. The short-range part is then formally defined as 
\begin{equation}
E^{\rm RPA, SR}_{\rm c} (\mu) = E^{\rm RPA}_{\rm c} - E^{\rm RPA, LR} (\mu),
\end{equation}
and is, in practice, approximated by a density functional
\begin{equation}
E_{\rm c}^{\rm RPA} \approx E^{\rm RPA,LR}_{\rm c} (\mu) + E^{\rm DFT, SR}_{\rm c} (\mu) .
\label{eq:rsRPA}
\end{equation}
This scheme provides a generalized adiabatic connection \cite{Toulouse2004,Yang1998} between the full-range RPA, which is obtained in the limit $\mu \to \infty$ and DFT in the limit $\mu \to 0$. As in standard DFT, the transferability of the density functional is key to the success of the method. Near the full-range limit, short-range effects are essentially localized. Thus, we can expect that the short-range part is described exactly by the local density approximation (LDA).\cite{Kohn1965} A more concrete theorem will be given further below. 

In the LDA, the short-range correction required to recover the full RPA correlation energy is given by
\begin{equation}
E^{\rm LDA, SR}_{\rm c} = \int \text{d}\textbf{r} \; \varepsilon^{\rm RPA, SR}_{\rm c, HEG}[n(\textbf{r})] n(\textbf{r}) , 
\label{eq:LDA}
\end{equation}
where $n(\textbf{r})$ is the local electronic density and $\varepsilon^{\rm RPA, SR}_{\rm c, HEG}$ is given by the difference  between the full-range and the long-range RPA correlation energy per particle for the homogeneous electron gas 
\begin{equation}
\varepsilon^{\rm RPA, SR}_{\rm c, HEG}(n) = \varepsilon_{\rm c, HEG}^{\rm RPA}(n)-\varepsilon^{\rm RPA,LR}_{\rm c, HEG}(n) .
\label{eq:short_range_LDA_def}
\end{equation}
Naturally, it is also possible to use more sophisticated density functionals. For example, various gradient correction schemes were discussed by \myciteauthor{Toulouse2005}.\cite{Toulouse2005} However, in this work we will stay at the LDA level of DFT, which was also the choice of \myciteauthor{Bruneval2012}. \cite{Bruneval2012}

Nowadays, the most common way to separate the Coulomb potential in long- and short-range parts is based on the error function
\begin{equation}
\begin{aligned}
&V^{\rm LR}(r,\mu) = \text{erf}(\mu r)/r  \\
&V^{\rm SR}(r,\mu) = 1/r-\text{erf}(\mu r)/r . 
\end{aligned}
\end{equation}
Besides the error function, other separation methods have been suggested. \citet{Toulouse2004} have used a scheme based on the ``erfgau interaction'' 
\begin{equation}\label{eq:erfgau}
\begin{aligned}
&V^{\rm LR}_{\rm erfgau}(r,\mu) = \text {erf}(c\mu r)/r - \frac{2c\mu}{\sqrt{\pi}}\exp(-c^2\mu^2r^2/3) ,\\
\end{aligned}
\end{equation}
where $c=(1+6\sqrt{3})^{1/2}$ is some scaling constant introduced in order to achieve similar cutoff radii as the error function. The erfgau interaction provides a much sharper separation between long and short-range interactions. They showed that this is manifest in the DFT limit, where both $E_{\rm x}^{\rm SR}$ and $E_{\rm c}^{\rm SR}$ were flat as a function of the range-separation parameter for the erfgau interaction, but not for the error function.\cite{Toulouse2004} This means that the erfgau interaction does a better job at separating out long-range effects. 

\subsection{Wave vector decomposition}

When working with a plane wave basis, it is natural to seek a range-separation scheme in Fourier space. In this
section, we review the available literature on what is commonly referred to as ``wave vector decomposition''.
We comment first on the limit of small wave vectors corresponding to the DFT limit (which is less relevant to the present work
but included here for completeness) and then on the more relevant case of large wave vectors, corresponding to the full-range limit.

The error function exhibits Gaussian decay over the Coulomb potential
\begin{equation}
V^{\rm LR}(q,\mu) = \frac{4\pi e^{-q^2/4\mu^2}}{q^2} ,
\label{eq:error_function}
\end{equation}
whereas the wave vector decomposition method is based on a hard momentum cutoff
\begin{equation}\label{eq:hardcutoff}
V^{\rm LR} (q,Q_{\rm cut}) =  \begin{cases} 4\pi/q^2 \hspace{15pt} & \text{for } q \leq Q_{\rm cut} \\
0 \hspace{15pt} & \text{for } q > Q_{\rm cut} .
\end{cases}
\end{equation}
The latter scheme was used by \citet{Nozieres1958} (NP) to investigate the RPA for the HEG.

For the small wave vector limit, NP developed a series expansion for $\varepsilon_{\rm c}^{\rm RPA,LR}$ in terms of the cutoff momentum
\begin{equation}
\begin{aligned}
& \varepsilon^{\rm RPA, LR}_{\rm c, HEG} (Q_{\rm cut}) =  -\frac{3\alpha}{8\pi}Q_{\rm cut}^2r_{\rm s} + \mathcal{O}(Q_{\rm cut}^3r_{\rm s}^{3/2}) ,
\label{NPexpansion}
\end{aligned}
\end{equation}  
where $\alpha = (4/9\pi)^{1/3}$ and $r_{\rm s}=(\alpha k_{\rm F})^{-1}$ is the Wigner-Seitz radius. NP studied corrections to the RPA from second and higher order exchange diagrams and found that they contribute to the expansion above only at order $\mathcal{O}(Q_{\rm cut}^4)$. Thus, the contribution of small wave vectors to the correlation energy is exactly described by the RPA
\begin{equation}
\varepsilon_{\rm c, HEG}^{\rm LR,RPA} \to  \varepsilon_{\rm c, HEG}^{\rm LR} \hspace{15pt} \text{for } Q_{\rm cut} \to 0 .
\end{equation}
The series expansion \eqref{NPexpansion} is valid, if the relevant momentum transfers are small compared to the Thomas-Fermi wave vector $k_{\rm s}=\sqrt{4k_{\rm F}/\pi}$.

For large $Q_{\rm cut}$, NP used second order perturbation theory for $\varepsilon_{\rm c, HEG}^{\rm SR}(Q_{\rm cut})$ and then interpolated between this and the small  $Q_{\rm cut}$ expansion \eqref{NPexpansion}. This gave the NP interpolation formula for the electronic correlation energy
\begin{equation}
\varepsilon_{\rm c, HEG}^{\rm NP} (r_{\rm s}) = \left( 0.031 \ln(r_{\rm s}) - 0.115\right) \text{Rydberg} . 
\end{equation}
A similar interpolation method was used by \citet{Langreth1975,Langreth1977} to describe the exchange-correlation energies of metallic surfaces. However, they used local density functional theory for large momentum transfers. Hence, their method closely resembles our own wave vector decomposition scheme.

\subsection{Approaching the full-range limit}

\begin{figure}[!!tb]
\centering
\includegraphics [width=\linewidth,keepaspectratio=true] {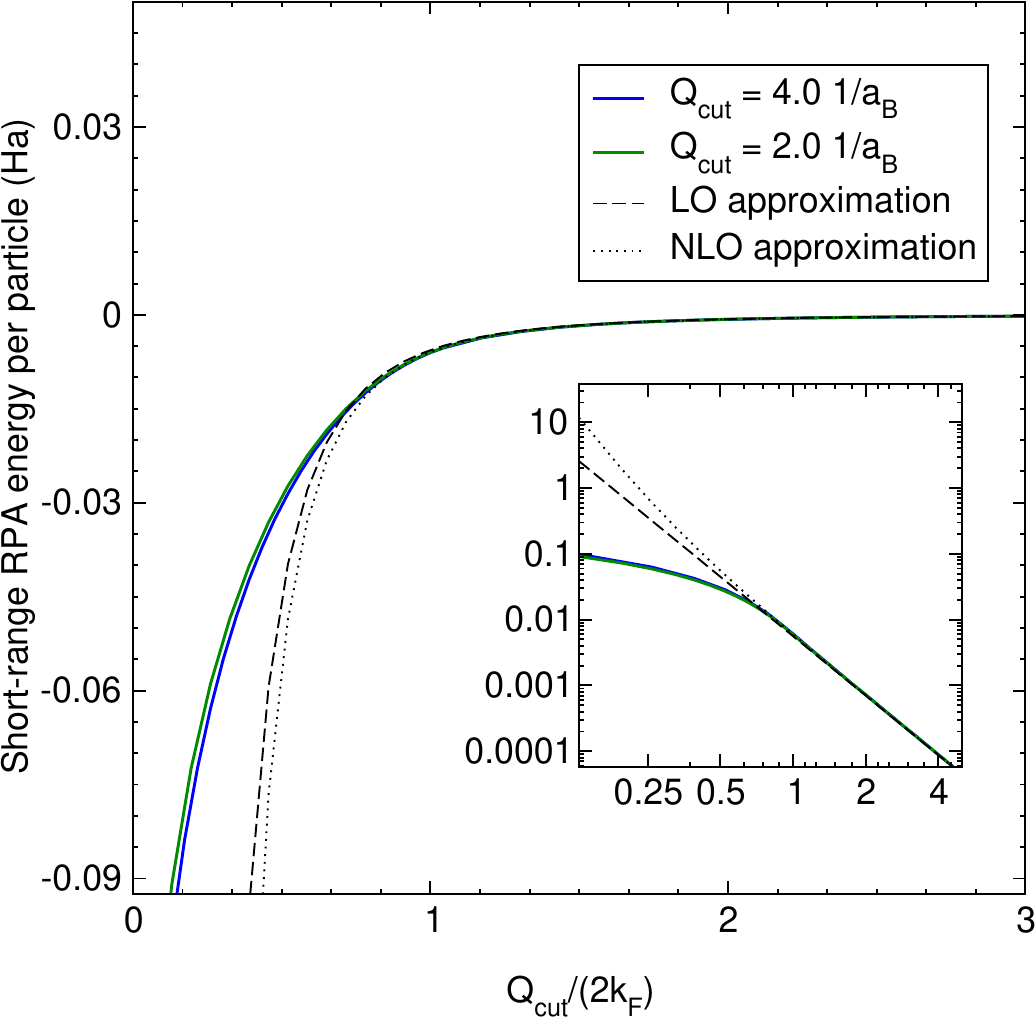}
\caption{Short-range LDA functionals for two different cutoffs $Q_{\rm cut}$. Full lines represent numerical evaluations of $\varepsilon_{\rm c, HEG}^{\rm RPA, SR}$ as a function of $1/k_{\rm F}\propto r_{\rm s}$, dashed and dotted lines represent the low density approximation \eqref{eq:Klimes_jellium} up to leading order (LO) and next-to-leading order (NLO) respectively. The latter is on this scale hardly distinguishable from $\varepsilon_{\rm c, HEG}^{\rm RPA, SR}$ for $Q_{\rm cut}/(2k_{\rm F}) \gtrsim 1$. Inset diagram: log-log plot of $|\varepsilon_{\rm c, HEG}^{\rm RPA,SR}|$.}
\label{fig:Gulans_expansion}
\end{figure}

The short-range RPA correlation energy per particle for the HEG, $\varepsilon_{\rm c, HEG}^{\rm RPA, SR}$, is an important quantity in our range-separation scheme [see Eq. \eqref{eq:LDA}]. One requires to know  $\varepsilon_{\rm c, HEG}^{\rm RPA, SR}$ as a function of the range-separation parameter $Q_{\rm cut}$. As  $Q_{\rm cut}\to \infty$, $V^{\rm LR}$ approaches the full Coulomb potential and $E_{\rm c}^{\rm RPA,SR}$ vanishes.
In Sec. \ref{sec:plane wave basis set incompleteness error}, we will show that the large $Q_{\rm cut}$ limit yields important information on the plane wave basis set incompleteness error. Furthermore, knowledge of the exact limits is useful for finding practical representations of $\varepsilon_{\rm c, HEG}^{\rm RPA, SR}$. \citet{Paziani2006} have conducted studies along these lines for the error function, here we concentrate on the hard cutoff.

Numerically, $\varepsilon_{\rm c, HEG}^{\rm RPA, SR}$ can be evaluated with relative ease, since the Lindhard response function for the HEG in terms of imaginary frequencies $\chi_{0, \rm HEG} (q,i\omega)$ is well known [see for example Eq. (4.8) in Ref. \onlinecite{vonBarth1972}]. For a given long-range potential and value of $r_{\rm s}$, we first obtain $\varepsilon_{\rm c,  HEG}^{\rm RPA, LR}(r_{\rm s})$ via (see Appendix \ref{App:Riemelmoser2020_B})
\begin{equation}
\begin{aligned}
& \varepsilon^{\rm RPA, LR}_{\rm c, HEG}
 = \frac{1}{n} \int \frac{\text{d}q}{(2\pi)^3}4\pi q^2 \int_{0}^{\infty}\frac{\text{d}\omega}{2\pi} \\
&\times \big[\ln \left(1-\chi_{0, \rm HEG}(q,i\omega)V^{\rm LR}(q) \right) \\
&+ \chi_{0, \rm HEG}(q,i\omega)V^{\rm LR}(q)\big] ,\\
\end{aligned}
\end{equation}
and then $\varepsilon_{\rm c, HEG}^{\rm RPA, SR}(r_{\rm s})$ using Eq.  \eqref{eq:short_range_LDA_def}.
We calculate $\varepsilon_{\rm c, HEG}^{\rm RPA, SR}$ according to this procedure and show results in Fig. \ref{fig:Gulans_expansion} for two different cutoffs $Q_{\rm cut}$.

In the low density or large $Q_{\rm cut}$ limit, $\varepsilon_{\rm c, HEG}^{\rm RPA,SR}$ behaves as (see Ref. \onlinecite{Gulans2014}, Appendix \ref{App:Riemelmoser2020_B})
\begin{equation}
\varepsilon^{\rm RPA, SR}_{\rm c, HEG} (r_{\rm s},Q_{\rm cut}) = - \frac{1}{\pi Q_{\rm cut}^3 r_{\rm s}^3} \hspace{3pt} + \mathcal{O}\left(\frac{1}{Q_{\rm cut}^5r_{\rm s}^5}\right).
\label{eq:Klimes_jellium}
\end{equation}
To estimate the validity range of the low density expansion, we have to compare $Q_{\rm cut}$ to some natural momentum scale. The relevant scale is the Fermi wave vector $k_{\rm F}$, which is manifest in Eq. \eqref{eq:Klimes_jellium}
($k_{\rm F}\propto 1/r_{\rm s}$). Fig. \ref{fig:Gulans_expansion} 
shows that the low density approximation is accurate for $Q_{\rm cut} \gtrsim 2k_{\rm F}$. Furthermore, Fig. \ref{fig:Gulans_expansion} shows that  $\varepsilon_{\rm c, HEG}^{\rm RPA, SR}$ is
nearly scale invariant when plotted versus $Q_{\rm cut}/ 2k_{\rm F}$. For $Q_{\rm cut}\to 0$, $\varepsilon_{\rm c, HEG}^{\rm RPA, SR}$ converges towards the full range RPA value $\varepsilon_{\rm c, HEG}^{\rm RPA}$, and for large $Q_{\rm cut}$, the scale invariance follows
from the asymptotics [Eq. \eqref{eq:Klimes_jellium}]. However, the near scale invariance at intermediate densities is not an obvious observation.  

Since we will also apply range-separation to the exchange energy, we require expressions
for the respective short-range LDA correction. Analogous to Eq. \eqref{eq:short_range_LDA_def}, the short-range exchange energy per particle for the HEG is defined as
\begin{equation} \label{eq:SR_exchange_analysis}
\varepsilon^{\rm SR}_{\rm x, HEG} (n) = \varepsilon_{\rm x, HEG}(n) - \varepsilon_{\rm x, HEG}^{\rm LR}(n),
\end{equation}
where $\varepsilon_{\rm x, HEG}=-3/(4\pi \alpha r_{\rm s})$ is the familiar Dirac expression for the full-range exchange energy per particle and $\varepsilon_{\rm x, HEG}^{\rm LR}$ is given by (see Appendix \ref{App:Riemelmoser2020_A})
\begin{equation} \label{eq:LR_exchange}
\begin{aligned}
&\varepsilon^{\rm LR}_{\rm x, HEG}(r_{\rm s},Q_{\rm cut})= \\
&\begin{cases}
\frac{-Q_{\rm cut}}{\pi} + \frac{3\alpha Q_{\rm cut}^2}{8\pi} r_{\rm s} - \frac{\alpha^3 Q_{\rm cut}^4}{64\pi} r_{\rm s}^3 \hspace{15pt} &\text{for } Q_{\rm cut} < 2k_{\rm F} \\
0 \hspace{15pt} &\text{for } Q_{\rm cut} \geq 2k_{\rm F} .
\end{cases}
\end{aligned}
\end{equation}
The constant term is connected to the normalization of the exchange hole, see Ref. \onlinecite{Yang1998}. As pointed out by NP, the term linear in $r_{\rm s}$ cancels the respective correlation term. This cancellation applies for the error function as well, but with linear terms $\pm 3\alpha/(2\pi) \mu^2r_{\rm s}$. \cite{Paziani2006}

\subsection{Plane wave basis set incompleteness error}\label{sec:plane wave basis set incompleteness error}

The large $Q_{\rm cut}$ limit of $\varepsilon^{\rm RPA, SR}_{\rm c, HEG}$ relates to an equation previously derived by \myciteauthor{Klimes2014a},\cite{Klimes2014a} as we briefly discuss in this section. To show this, one inserts the leading order term in Eq. (\ref{eq:Klimes_jellium}) into
Eq. (\ref{eq:LDA}) and uses $r_{\rm s}=(3 /( 4 \pi n) )^{1/3}$. This means we construct a local density functional approximation for the short-range part, yielding
\begin{equation} \label{eq:SR_LDA_theorem}
 E_{\rm c}^{\rm RPA,SR}(Q_{\rm cut}) \approx - \frac{4}{3Q_{\rm cut}^3} \int \text{d}\textbf{r} \, n(\textbf{r})n (\textbf{r}) . 
\end{equation}
By Fourier transforming the density to reciprocal space, we obtain
\begin{equation}
\ E_{\rm c}^{\rm RPA,SR}(Q_{\rm cut}) \approx - \frac{4\Omega}{3 Q_{\rm cut}^3} \sum_{\textbf{q}} |n(\textbf{q})|^2 , 
\label{eq:Klimes_full}
\end{equation}
where $\Omega$ is the system volume. This is exactly the expression that has been derived by \citet{Klimes2014a} for the plane wave basis set incompleteness error [their definition of the latter, compare Eq. (10) in Ref. \onlinecite{Klimes2014a}, corresponds to $E^{\rm RPA, SR}_{\rm c}$ in our range-separation scheme].

This shows that at sufficiently high cutoffs the plane wave basis set incompleteness error is described exactly by the short-range LDA. It is important to point out that this statement contains additional information over the simple fact that $E^{\rm RPA, LR}_{\rm c}(Q_{\rm cut})$ approaches the full-range value as $Q_{\rm cut} \to \infty$, since it precisely predicts the leading order correction. 

The key assumption in the derivation of \citet{Klimes2014a} was that for high energies the unoccupied orbitals can be approximated by plane waves. This is always guaranteed as the kinetic energy then dominates the Hamiltonian. Clearly, using the HEG as a reference system makes a similar assumption. An alternative proof of this theorem using the coupling constant formalism was given previously by \myciteauthor{Burke1994}.\cite{Burke1994} \citet{Toulouse2004} have derived similar theorems for the error function. 

Finally, we briefly discuss how these results can be extended beyond the RPA. In the RPA+SOSEX (second order screened exchange),\cite{Grueneis2009,Paier2010,*Paier2010a} Eq. \eqref{eq:SR_LDA_theorem}	is merely reduced by a factor of two for non-spin-polarized systems. This can be easily understood, as the SOSEX correction simply restores the Pauli principle in the large $Q_{\rm cut}$ limit, i.e. the self-interaction between electrons of same spins is exactly cancelled.\cite{Nozieres1958,Maggio2016} However, the exact (rather than RPA) short-range correlation energy is not simply a function of the local density alone, but also involves the on-top pair density. \cite{Burke1994,Toulouse2004} Although the LDA does not generally describe the latter exactly, it is still an accurate approximation, if the ground state wave function in the weak coupling limit is well described by a single Slater determinant. \cite{Burke1998}

\subsection{Plane wave basis set extrapolation}

\begin{figure}[!!tb]
\centering
\includegraphics [width=\linewidth,keepaspectratio=true] {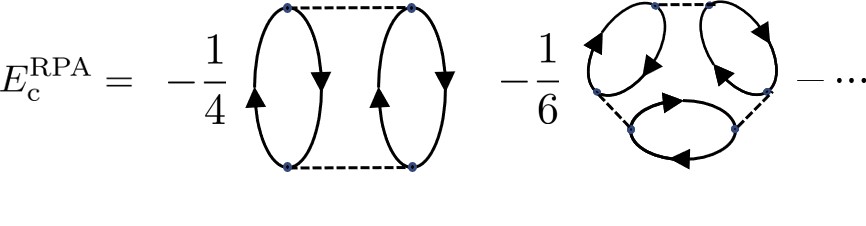}
\caption{Representation of the RPA correlation energy in terms of Feynman diagrams. The Feynman rules can be found in Ref. \onlinecite{Fetter2003}, with prefactor convention as in Ref. \onlinecite{Ren2012}. The second order ring diagram is called direct MP2 diagram.}
\label{fig:diagrams}
\end{figure}

Prior to the work of Klime\ifmmode \check{s}\else \v{s}\fi{} \textit{et al.},\cite{Klimes2014a} \citet{Harl2008} provided numerical evidence that the plane wave basis set incompleteness error falls off as $1/Q_{\rm cut}^3$ and suggested a basis set extrapolation method based on
\begin{equation}
E^{\rm RPA, LR}_{\rm c}(Q_{\rm cut}) = E^{\rm RPA}_{\rm c} + \frac{\rm A_3}{Q_{\rm cut}^3} ,
\label{eq:Harl_extrapolation}
\end{equation}
where $E^{\rm RPA}_{\rm c}$ and $A_3$ are obtained by a linear fit of $E^{\rm RPA, LR}_{\rm c} (Q_{\rm cut})$ versus $1/Q_{\rm cut}^3$.
\citet{Gulans2014} has calculated higher order corrections to Eq. \eqref{eq:Klimes_jellium} and suggested the extrapolation
\begin{equation}
\begin{aligned}
E^{\rm RPA, LR}(Q_{\rm cut}) = & E^{\rm RPA}_{\rm c} + \\
& \frac{\rm A_3}{Q_{\rm cut}^3} + \frac{\rm A_5}{Q_{\rm cut}^5} + \frac{\rm A_7}{Q_{\rm cut}^7},
\end{aligned}
\end{equation}
where $A_5$ and $A_7$ are further fit parameters. In Appendix \ref{App:Riemelmoser2020_B}, we provide an analytical evaluation of $A_3$ - $A_7$ for the HEG, yielding

\begin{equation}
\label{eq:Gulans_expansion}
\begin{aligned}
& E^{\rm RPA, SR}_{\rm c, HEG} (Q_{\rm cut})/\Omega = \\
-&n^2 \left[ \frac{4}{3Q_{\rm cut}^3} + \frac{8k_{\rm F}^2}{25Q_{\rm cut}^5} + \frac{288k_{\rm F}^4}{1225Q_{\rm cut}^7} + \mathcal{O}\left(\frac{1}{Q_{\rm cut}^9}\right) \right] \\
+&\pi n^3 \left[ \frac{32}{7Q_{\rm cut}^7} + \mathcal{O}\left(\frac{1}{Q_{\rm cut}^9}\right)\right] + ... ,\\ 
\end{aligned}
\end{equation}
where $n$ is the electronic density of the HEG [note that our coefficients differ from the ones given in Ref. \onlinecite{Gulans2014}, see discussion after Eq. \eqref{eq:Giuliani}]. The terms proportional to $n^2$ stem from the direct MP2 diagram, compare Fig. \ref{fig:diagrams}. The terms proportional to $n^3$ stem from the third order ring diagram and so forth. As was pointed out by Gulans, these terms can be interpreted neatly in terms of the electronic cusp. They represent probabilities that there are two, three, ..., electrons at the same place. This interpretation suggests a beyond-LDA short-range functional based on an expansion involving the pair density. However, we do not attempt to construct such a functional here and leave it up to future work. 

\subsection{One-shot extrapolation method}

The basis set extrapolation schemes described above require repeated evaluations of RPA correlation energies for a set of cutoff energies $Q_{\rm cut}$. Since this is typically one of the bottlenecks in modern RPA implementations, the basis set extrapolation increases the computational cost. We now propose a one-shot extrapolation scheme that accounts for basis set incompleteness \textit{a priori} via an optimized long-range potential. The Coulomb interaction is enhanced at intermediate momentum transfers $q \approx Q_{\rm cut}$ in order to make up for the cutoff at $q = Q_{\rm cut}+\Delta Q$
\begin{equation}
\begin{aligned}
&V^{\rm LR}_{\rm SCK}(q,Q_{\rm cut}) =   \\
&\begin{cases} 4\pi/q^2 &\text{for } q < Q_{\rm cut}-\Delta Q \\
4\pi f_{\rm SCK}(q) /q^2 &\text{for } Q_{\rm cut}-\Delta Q \leq q  \leq Q_{\rm cut}+\Delta Q \\
0 &\text{for } q > Q_{\rm cut}+\Delta Q .
\end{cases}
\end{aligned}
\end{equation}
We base the construction of this ``squeezed Coulomb kernel'' (SCK) on the fact that at large momentum transfers, the RPA is dominated by the direct MP2 term (see Fig. \ref{fig:diagrams})
\begin{equation}
\varepsilon^{\rm RPA}_{\rm c, HEG}(q)  \sim \chi_{0, \rm HEG}^2(q)V^2(q) \hspace{15pt} \text{for } q \to \infty ,
\end{equation}
 where $\chi_{0, \rm HEG}(q)$ is the frequency integrated Lindhard response function, see Appendix \ref{App:Riemelmoser2020_B}. In the same limit, $\chi_{0, \rm HEG}(q)$ falls off as $1/q$, which yields the leading order term in Eq. \eqref{eq:Gulans_expansion}
\begin{equation}
\varepsilon^{\rm RPA,SR}_{\rm c, HEG}(Q_{\rm cut}) \sim \int_{Q_{\rm cut}}^{\infty} \text{d}q \; q^2 \left(\frac{1}{q} V(q)\right)^2 \sim \frac{1}{Q_{\rm cut}^3} .
\end{equation}
The SCK is constructed to make up for the loss of correlation energy by enhancing the Coulomb potential at intermediate $q$ assuming $\chi_0 \approx \chi_{0, \rm HEG}$
\begin{equation}
\label{eq:Attenuated_kernel}
\begin{aligned}
&\int_{Q_{\rm cut}-\Delta Q}^{\infty}\text{d}q \; q^2\left[ \frac{1}{q} V(q) \right]^2 \\
& \overset{!}{=} \int_{Q_{\rm cut}-\Delta Q}^{\infty} \text{d}q \; q^2\left[ \frac{1}{q} V^{\rm LR}_{\rm SCK}(q,Q_{\rm cut}) \right]^2 \\
&= \int_{Q_{\rm cut}-\Delta Q}^{Q_{\rm cut}+\Delta Q} \text{d}q \; q^2\left[ \frac{4\pi}{q^3} f_{\rm SCK}(q) \right]^2 .
\end{aligned}
\end{equation}
Details on the choice of $f_{\rm SCK}(q)$ and $\Delta Q$ will be given in section \ref{sec:Riemelmoser2020_3}, where we also show that $\Delta Q$ plays only the role of a window parameter. This means that the effective cutoff momentum is located at $q=Q_{\rm cut}$, i.e. at the center of the window, rather than at $q=Q_{\rm cut}-\Delta Q$.

This approach is in spirit similar to standard ion-pseudo-potential methods, which have been employed to handle the nuclear cusp. \cite{Heine1970} The  standard pseudo-potentials are constructed such that they reproduce the electronic properties of an atomic reference. In our case, it is the low-density HEG that plays the role of the reference system. 

We do not combine the SCK with a short-range LDA correction, because the SCK already describes low densities well enough. This is exactly the region where the LDA works best, while for high densities it transfers spurious long-range effects to inhomogeneous systems.\cite{Toulouse2005,Langreth1983,*Langreth1984} 

In fact, we argue in the following that neglecting the LDA correction can be seen as an attempted effective gradient correction. In the local interaction parameter effective gradient correction of \myciteauthor{Toulouse2005},\cite{Toulouse2005} $\varepsilon_{\rm c, HEG}^{\rm SR}$ is reduced for small $r_{\rm s}$ to prevent the over-correction of correlation. This was done by choosing the range-separation parameter locally as
\begin{equation}
\mu_{\rm eff} = \max[\mu_{\rm l}(\textbf{r}),\mu] ,
\end{equation} 
where $\mu_{\rm l} (\textbf{r})$ is some typical correlation length, for instance $\mu_{\rm l}(\textbf{r})=\alpha k_{\rm s}(\textbf{r})$. This reduces the short-range correction for high densities and hence mimics a gradient correction, as
\begin{equation}
\varepsilon^{\rm RPA,SR}_{\rm c, HEG}(r_{\rm s},\mu_1) < \varepsilon^{\rm RPA,SR}_{\rm c, HEG}(r_{\rm s},\mu_2) \hspace{15pt} \text{if } \mu_1 > \mu_2 .
\end{equation}
We return now to the SCK and assume that $\varepsilon^{\rm RPA,SR}_{\rm c, HEG}$ vanishes exactly for $Q_{\rm cut}>\mu_{\rm l}$. We will show in section \ref{sec:Riemelmoser2020_3} that this is an excellent approximation for $\mu_{\rm l} = 2k_{\rm F}$. Then, an effective gradient correction is obtained by simply neglecting the short-range LDA correction, as the assumption above implies

\begin{equation}
\varepsilon^{\rm RPA,SR}_{\rm c, HEG, SCK}(r_{\rm s},\max[\mu_{\rm l},Q_{\rm cut}]) = 0 \hspace{15pt} \forall r_{\rm s} .
\end{equation}

\section{\label{sec:Riemelmoser2020_3} Computational details}

In the following we present results from various numerical studies of the HEG. We argue that for the purpose of basis set correction, one can identify $\mu \approx Q_{\rm cut}$ for a fair comparison between the error function and the plane wave cutoff schemes. Since its structure is visually more revealing, we show the long-range RPA correlation energy per particle $\varepsilon_{\rm c, HEG}^{\rm RPA, LR}$ rather than $\varepsilon_{\rm c, HEG}^{\rm RPA, SR}$ throughout. The latter can be easily obtained by subtracting $\varepsilon_{\rm c, HEG}^{\rm RPA, LR}$ from the full-range reference. 

\subsection{Smooth momentum cutoff}

\begin{figure}[!!tb]
\centering
\includegraphics [angle=-90,width=\linewidth,keepaspectratio=true] {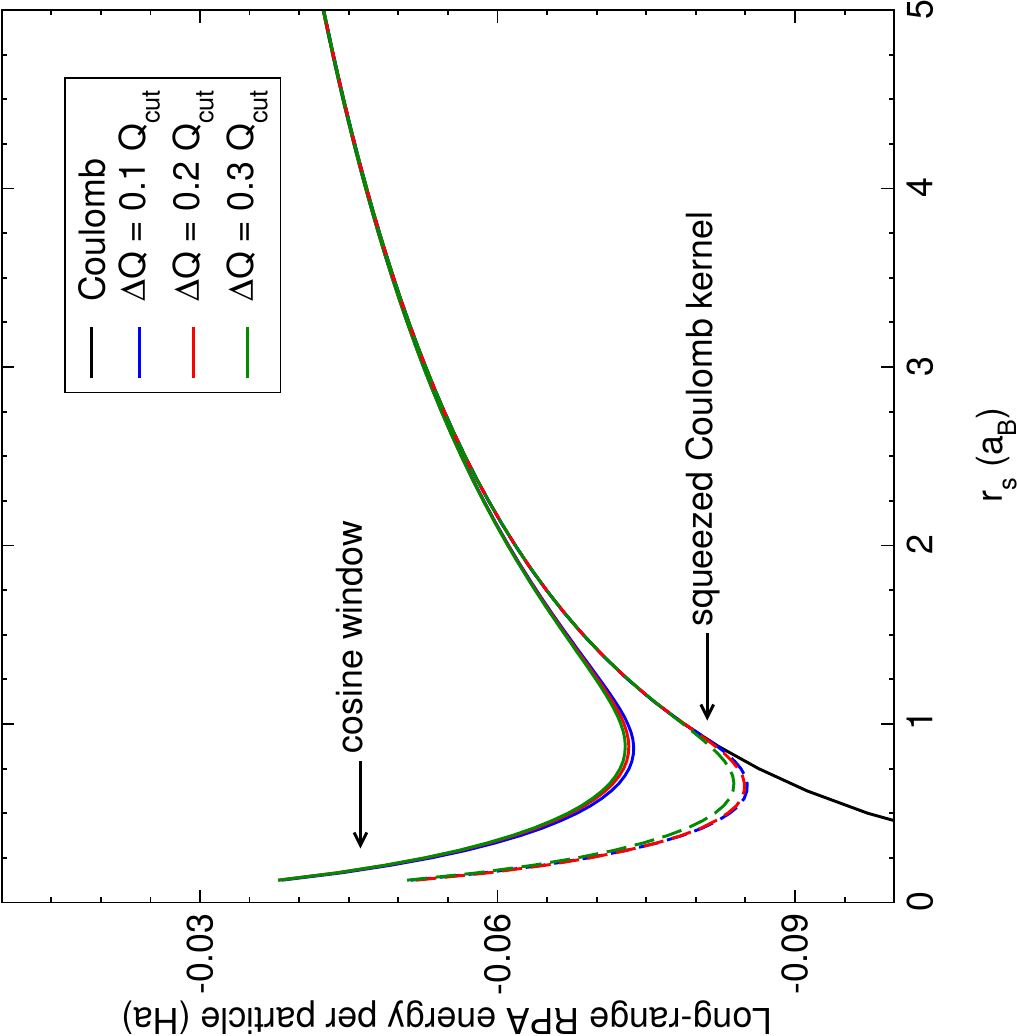}
\caption{Long-range RPA correlation energies per particle for the HEG, $\varepsilon^{\rm RPA,LR}_{\rm c, HEG}$, for different window parameters $\Delta Q$ at fixed cutoff momentum $Q_{\rm cut} = 4.0 \: a_{\rm B}^{-1}$. Solid lines represent the cosine window, dashed lines the squeezed Coulomb kernel (SCK). }
\label{fig:windows}
\end{figure}

The discontinuous cutoff in Eq. \eqref{eq:hardcutoff} causes technical problems such as shell-filling effects. We follow Ref. \onlinecite{Harl2010} and replace the step at $q=Q_{\rm cut}$ by a smooth cosine window 
\begin{equation}
\begin{aligned}
&V^{\rm LR}_{\rm cos}(q,Q_{\rm cut}) =   \\
&\begin{cases} 4\pi/q^2 &\text{for } q < Q_{\rm cut}-\Delta Q \\
4\pi f_{\rm cos}(q) /q^2 &\text{for } Q_{\rm cut}-\Delta Q \leq q  \leq Q_{\rm cut}+\Delta Q \\
0 &\text{for } q > Q_{\rm cut}+\Delta Q ,
\end{cases}
\end{aligned}
\end{equation}
where 
\begin{equation}
\begin{aligned}
& f_{\rm cos}(q) = \frac{1}{2}\\
+&\frac{1}{2} \cos\left[\frac{q^2/2 - (Q_{\rm cut}-\Delta Q)^2/2}{(Q_{\rm cut}+\Delta Q)^2/2-(Q_{\rm cut}-\Delta Q)^2/2}\pi\right] ,
\end{aligned}
\end{equation} 
and $\Delta Q$ controls the ``smoothness'' of the window. For $\Delta Q \to 0$, Eq. \eqref{eq:hardcutoff} is recovered. Even a narrow window with $\Delta Q = 0.1Q_{\rm cut}$ remedies the technical issues, and changes the short-range functional very little. Most notable is a slight reduction of $\varepsilon^{\rm RPA, LR}_{\rm c, HEG}$ at densities where $r_{\rm s} \approx 4/Q_{\rm cut}$ [see Fig. \ref{fig:windows} (solid lines)]. The choice of the window parameter $\Delta Q$ represents a trade-off between smoothness and plane wave basis set convergence, since the latter is now tied to a cutoff energy $(Q_{\rm cut} + \Delta Q)^2/2$. 

\subsection{Form of the squeezed Coulomb kernel}

\begin{figure}[!!tb]
\centering
\includegraphics [width=\linewidth,keepaspectratio=true] {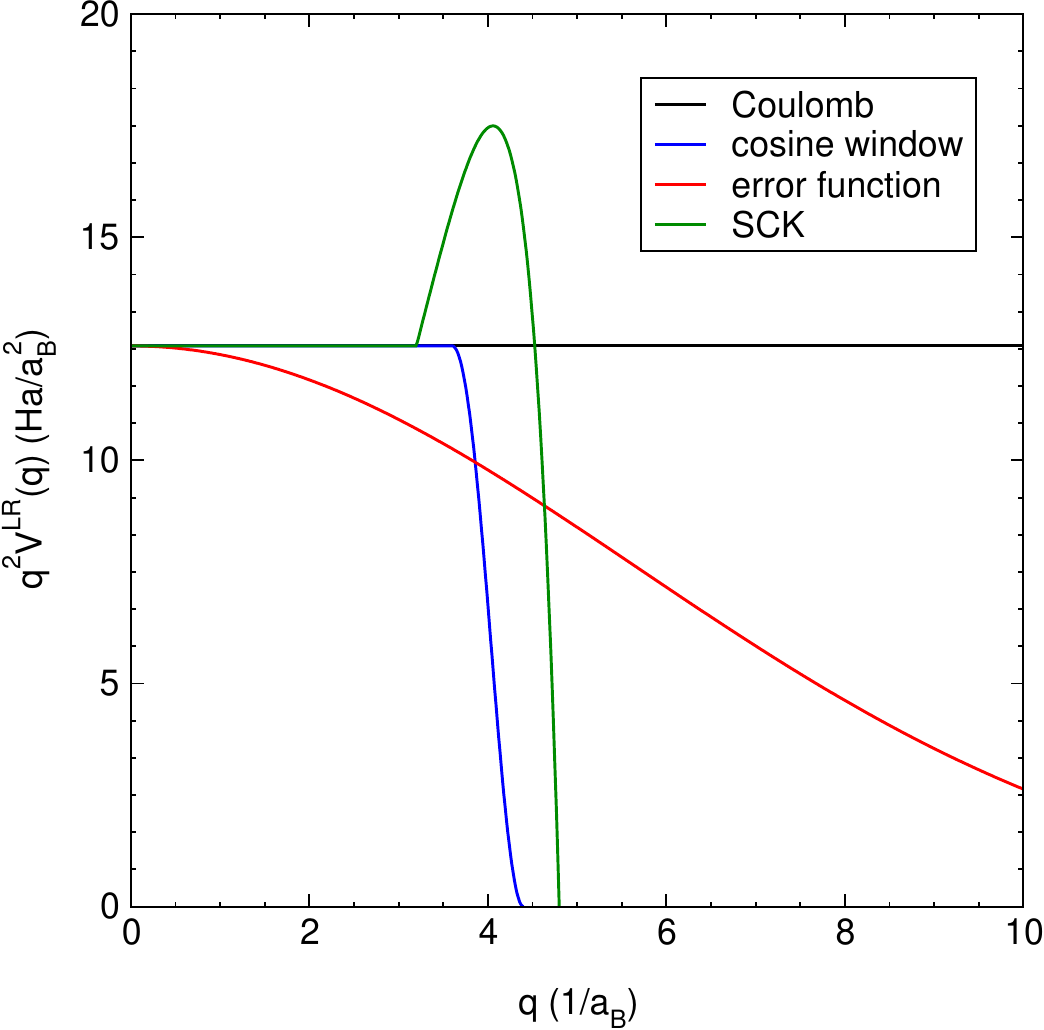}
\caption{Fourier representation of the different long-range potentials at range-separation parameters $Q_{\rm cut}=\mu = 4.0 \: a_{\rm B}^{-1}$. For the width of the cosine window, we use $\Delta Q = 0.1 Q_{\rm cut}$, and for the squeezed Coulomb kernel (SCK) $\Delta Q=0.2Q_{\rm cut}$.}
\label{fig:LR_potentials}
\end{figure}

In the construction of the SCK, we impose the following constraints: (i) it should join onto the bare Coulomb kernel at $Q_{\rm cut}-\Delta Q$, (ii) it should vanish at $Q_{\rm cut} + \Delta Q$, and (iii) it should be positive definite. A convenient form that satisfies the constraints (i)-(iii) as well as Eq. \eqref{eq:Attenuated_kernel} is
\begin{equation}
f_{\rm SCK}(q) = q^2 \frac{2\Delta Q(Q_{\rm cut}+\Delta Q-q)}{[(Q_{\rm cut}-\Delta Q)^2-q(Q_{\rm cut}-3\Delta Q)]^2} .
\end{equation}
Similar to the smooth momentum cutoff above, $\Delta Q$ controls the width of the window [see Fig. \ref{fig:windows} (dashed lines)]. As the function $f_{\rm SCK}(q)$ contains more structure than the cosine window, we use a broader window with $\Delta Q=0.2 Q_{\rm cut}$. The form of the SCK is displayed in Fig. \ref{fig:LR_potentials}. As variations in the window parameter $\Delta Q$ do not change $\varepsilon_{\rm c, HEG}^{\rm RPA, LR}$ significantly, the effective cutoff momentum is located at the center of the window, i.e. at $q=Q_{\rm cut}$, as is the case for the cosine window as well.

\subsection{Analytical representation of the short-range local density functional}

\begin{figure}[!!tb]
\centering
\subfloat{\includegraphics [width=\linewidth,keepaspectratio=true] {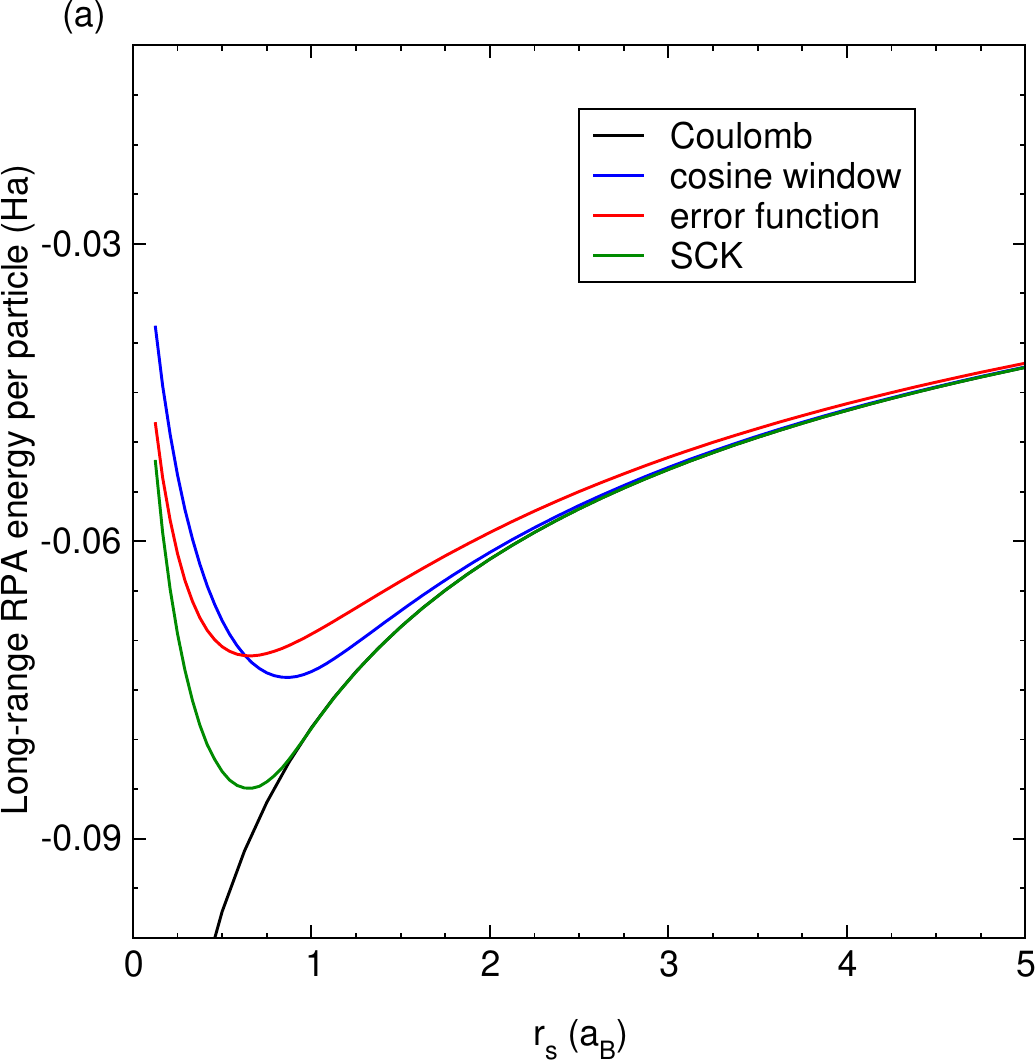}\label{fig:LR_energies_a}}
\par\medskip
\subfloat{\includegraphics [width=\linewidth,keepaspectratio=true] {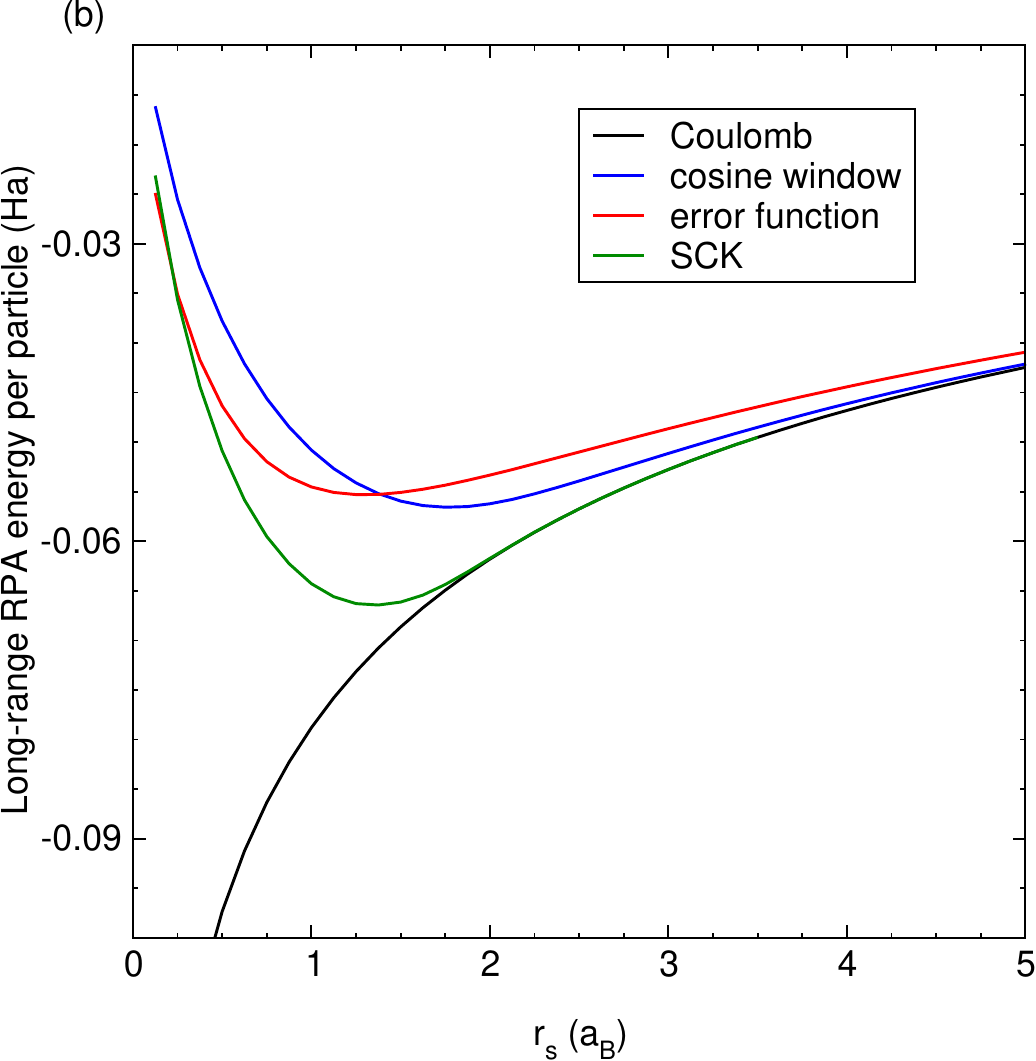}\label{fig:LR_energies_b}}
\caption{Long-range RPA correlation energies per particle $\varepsilon^{\rm RPA, LR}_{ \rm c, HEG}$ for the HEG for the different long-range potentials. \protect\subref{fig:LR_energies_a} range-separation parameters $Q_{\rm cut}=\mu = 4.0 \: a_{\rm B}^{-1}$ as in Fig. \ref{fig:LR_potentials}, \protect\subref{fig:LR_energies_b} $Q_{\rm cut}=\mu = 2.0 \: a_{\rm B}^{-1}$.}
\label{fig:LR_energies}
\end{figure}

Fig. \ref{fig:LR_energies} depicts $\varepsilon^{\rm RPA, LR}_{\rm c, HEG}$ for different long-range potentials. Generally, $\varepsilon_{\rm c, HEG}^{\rm RPA, LR}$ vanishes for $r_{\rm s} \to 0$ and approaches the full-range value as $r_{\rm s} \to \infty$, compare Eqs. \eqref{NPexpansion} and \eqref{eq:Klimes_jellium}. For each long-range potential, there is a distinct minimum at intermediate densities, which indicates a shift from the high density to the low density regime. The relevant limit for basis set correction is that of large range-separation parameters, corresponding to the low density regime. Judging from the numerical data, we identify this regime roughly at densities given by $r_{\rm s} \gtrsim 4/Q_{\rm cut}$ and $r_{\rm s} \gtrsim 4/\mu$, which corresponds to $Q_{\rm cut}\gtrsim 2k_{\rm F}$ and $\mu \gtrsim 2k_{\rm F}$.

For the error function, $\varepsilon^{\rm RPA, LR}_{\rm c, HEG}$ approaches the full-range value only slowly as $r_{\rm s} \to \infty$. Identifying $Q_{\rm cut}\approx\mu$, the cosine window describes low densities better than the error function, but high densities worse. This means that the cosine window separates long- and short-range effects better than the error function. The SCK is exact in the low density limit per construction, and $\varepsilon^{\rm RPA,SR}_{\rm c, HEG} = \varepsilon_{\rm c, HEG}^{\rm RPA} - \varepsilon^{\rm RPA, LR}_{\rm c, HEG}$  vanishes rapidly for $Q_{\rm cut}\gtrsim 2k_{\rm F}$.

To parametrize $\varepsilon^{\rm RPA, SR}_{\rm c, HEG}(r_{\rm s})$ for both the cosine window and the error function, we choose a form similar to that used by \citet{Paziani2006}
\begin{equation}
\varepsilon^{\rm RPA, SR}_{\rm c, HEG} (r_{\rm s}) =A \frac{\ln\left( \frac{r_{\rm s} + a_0 r_{\rm s}^2 +a_1 r_{\rm s}^3 + a_2 r_{\rm s}^4 }{1 + a_3 r_{\rm s} + a_4 r_{\rm s}^2 +a_5 r_{\rm s}^3 + a_2 r_{\rm s}^4}\right)}{1+a_6 r_{\rm s} + a_7 r_{\rm s}^2}.
\label{eq:Paziani_fit}
\end{equation}
The $a_i$ are determined by a non-linear least square fit for a given range-separation parameter and the constant $A$ enforces the high density limit of \citet{Gell-Mann1957} 
\begin{equation}
\begin{aligned}
& \varepsilon^{\rm RPA, SR}_{\rm c, HEG} (r_{\rm s}) \approx A \ln(r_{\rm s}) \hspace{15 pt} \text{for } r_{\rm s} \to 0 \\
& A = -\frac{\ln(2)-1}{\pi^2} .
\end{aligned}
\end{equation} 
This is the appropriate behavior for any reasonable LDA functional, since (i) $\varepsilon_{\rm c, HEG}^{\rm RPA,LR}$ vanishes for $r_{\rm s} \to 0$, and (ii) the RPA becomes exact in this limit. The fit parameters $a_i$ are given in Appendix \ref{App:Riemelmoser2020_C} for selected range-separation parameters. 

\section{\label{sec:Riemelmoser2020_4} Applications}

\subsection{Applied settings}

\begin{table}[!tb] 
\begin{ruledtabular}
\caption{Chosen settings for the RPA calculations. The short-hand \_sv for the PAW potentials indicates treatment of the outermost $s$ and $p$ core states as valence. The orbital plane wave cutoffs $E_{\rm max}$ are given in eV, $k$-point sampling is done on $\Gamma$-centered $k\times k \times k$-grids. Values for the zero point corrected experimental lattice constant $a_0$ (in \AA) are taken from Ref. \onlinecite{Schimka2011}. For Kr, we use the result from an accurate quantum chemistry calculation. \cite{Rosciszewski2000} Values of $2k_{\rm F}$ at the experimental lattice constant are given in $a_{\rm B}^{-1}$. }
\label{tab:settings}
\begin{tabular}{lcccccc}
 &  & PAW potentials & $E_{\rm max}$ & $k$ & $a_0$ & $2k_{\rm F}$\\ 
\hline \\\\[-3.\medskipamount]
C & (A4)  & C\_GW\_new & 600 & 6 & 3.553 & 2.93\\ 
Si & (A4) & Si\_GW & 500 & 6 & 5.421 & 1.92\\
MgO & (B1) & Mg\_sv\_GW O\_GW & 600 & 6 & 4.189 & 3.13\\
Kr & (A1)  & Kr\_GW & 700 & 8 & 5.598 & 1.88\\
Pd & (A1)   & Pd\_sv\_GW & 500 & 10 & 3.876 & 3.51\\
Al & (A1) & Al\_sv\_GW & 600 & 10 & 4.018 & 2.88 
\end{tabular}
\end{ruledtabular}
\end{table}

We use the RPA implementation in the Vienna Ab Initio Simulation Package (VASP) \cite{Kresse1996} as presented in Ref. \onlinecite{Harl2010}, which also describes the calculation of the exact exchange energy. The orbitals of the underlying DFT calculations are obtained self-consistently using the Perdew-Burke-Ernzerhof (PBE) functional. \cite{Perdew1996,*Perdew1997} The equilibrium lattice constants are found by a Murnaghan equation of state fitted to energies evaluated at seven lattice volumes centered at the experimental value, spanning a window of $\pm 15\%$ . If the lattice constants deviate from experiment by more than 3\%, additional points are considered. In the present work we consider six prototypical materials, which are summarized in Table \ref{tab:settings}. Notably, we use a large plane wave cutoff energy $E_{\rm max}$ for the orbital basis set  (\texttt{ENCUT} in VASP). This ensures sufficiently converged lattice constants for all long-range potentials. The relative errors in the lattice constants are estimated to be under 0.1\%. For the response function  $\chi_0$, we use a smaller cutoff energy $E_{\rm max}^{\chi}=2/3 \: E_{\rm max}$  (\texttt{ENCUTGW} in VASP). 

\subsection{Convergence behavior for total energies}

\begin{figure}[!!tb]
\centering
\subfloat{\includegraphics [angle=-90,width=\linewidth,keepaspectratio=true] {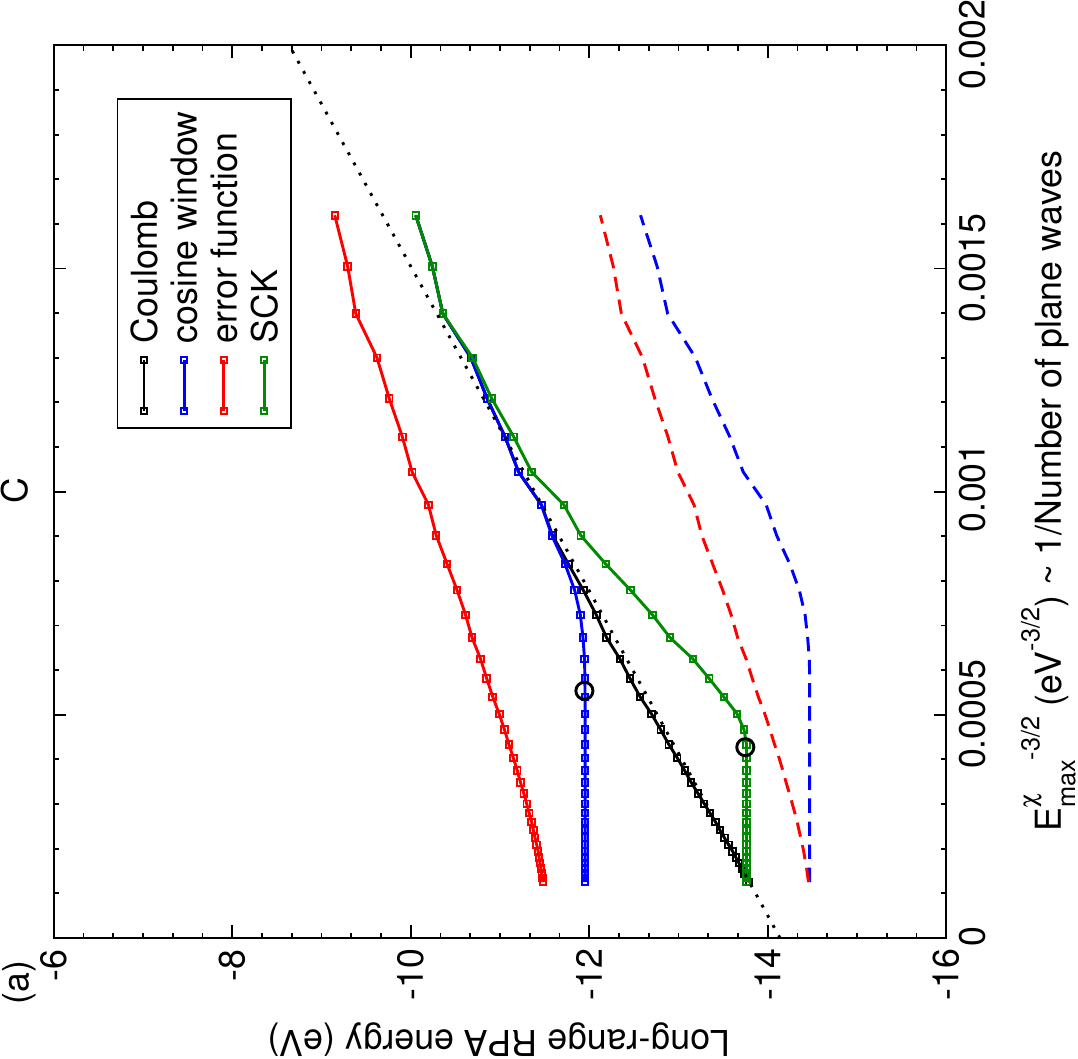}\label{fig:convergence_a}}
\par\medskip
\subfloat{\includegraphics [angle=-90,width=\linewidth,keepaspectratio=true] {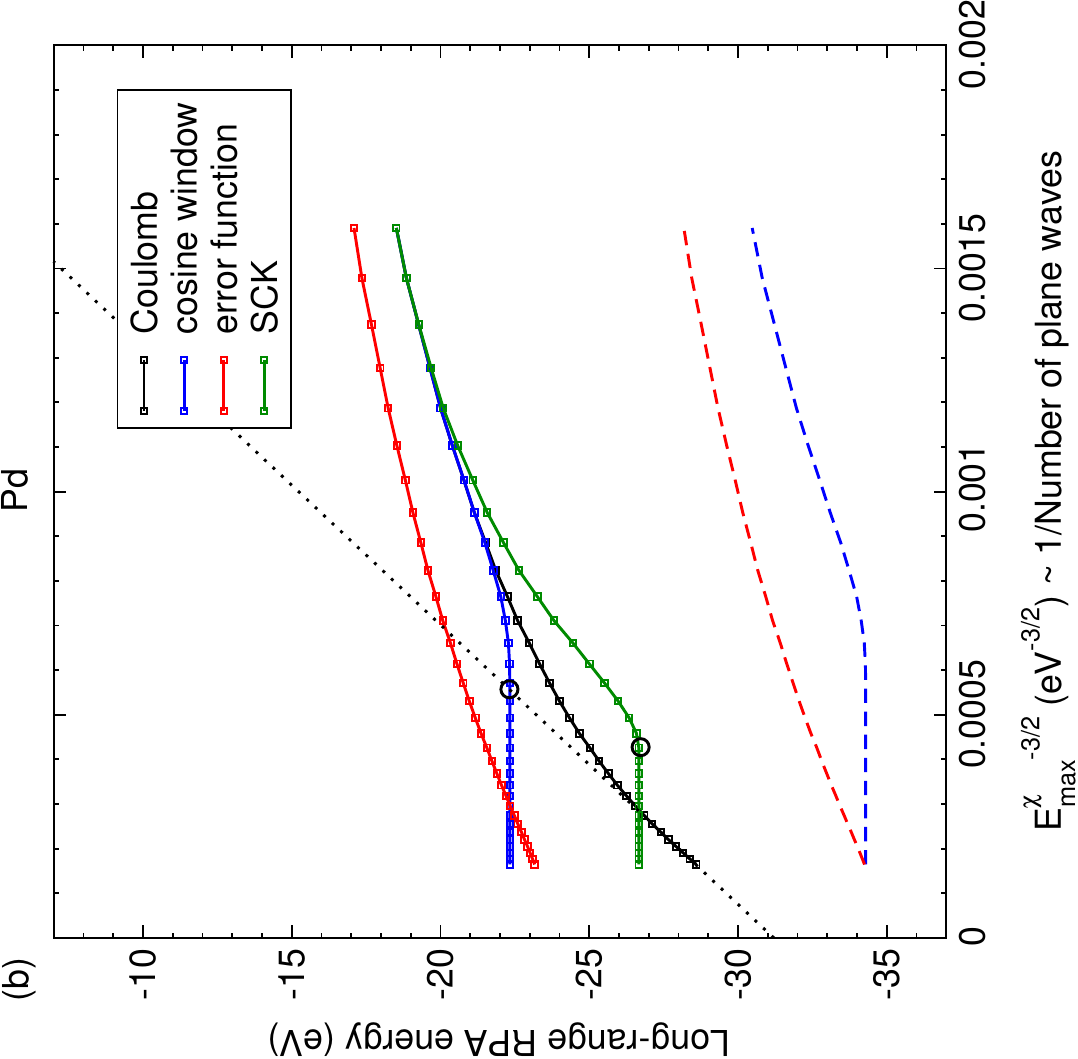}\label{fig:convergence_b}}
\caption{Convergence of $E^{\rm RPA,LR}_{\rm c}$ with respect to $E_{\rm max}^{\chi}$ for \protect\subref{fig:convergence_a} C and \protect\subref{fig:convergence_b} Pd at the experimental lattice volume. Range-separation parameters are set to $Q_{\rm cut}=\mu = 3.0 \: a_{\rm B}^{-1}$. Dotted lines represent basis set extrapolations using Eq. \eqref{eq:Harl_extrapolation}, fitted to the last eight data points (linear fits in this plot). For cosine window and error function, dashed lines include the short-range LDA correction. Circles indicate the energies $(Q_{\rm cut}+\Delta Q)^2/2$, where the long-range potentials vanish for the cosine window and the SCK respectively.}
\label{fig:convergence}
\end{figure}

Fig. \ref{fig:convergence} compares the convergence behavior of the long-range RPA energies with respect to $E^{\chi}_{\rm max}$ for C and Pd at the experimental lattice constant, with $E_{\rm max}$ fixed to the values stated in Table \ref{tab:settings}. For both the cosine window and the SCK, $E^{\rm RPA, LR}$ is fully converged with respect to $E_{\rm max}^{\chi}$ once the potentials vanish, i.e. beyond the energies $(Q_{\rm cut}+\Delta Q)^2/2$. These cutoff energies are given in Table \ref{tab:cutoffs} for selected range-separation parameters. 

This makes it possible to automatically adapt the range-separation parameter $Q_{\rm cut}$ to a given plane wave cutoff $E^{\chi}_{\rm max}$ or vice versa. For the error function, however, $E^{\rm RPA, LR}_{\rm c}$ is in principle fully converged only at infinite cutoff energies, and this simple automatic adaption is not possible. In practice, one would have to perform additional convergence tests for each value of $\mu$ or resort to more elaborate schemes.\cite{Loos2019} For the sake of simplicity, we omit these convergence tests throughout and use the large cutoffs $E_{\rm max}^{\chi}=2/3 \: E_{\rm max}$ for all values of $\mu$, compare Table \ref{tab:settings}.

Fig. \ref{fig:convergence} shows that replacing $E^{\rm RPA}_{\rm c} \rightarrow E^{\rm RPA, LR}_{\rm c}$ generally underestimates the correlation energy, whereas including the short-range LDA correction (dashed curves) overcorrelates. Fig. \ref{fig:convergence} furthermore demonstrates the basis set extrapolation according to Eq. \eqref{eq:Harl_extrapolation} (dotted curves, $1/N_{\rm PW} \propto 1/Q_{\rm cut}^3$), which provides full-range reference values in the following. 

For C, the full-range data closely follows the $1/N_{\rm PW}$ behavior that is expected from the HEG model even at smaller values of ${E_{\rm max}^{\chi}}$. In contrast, the full-range curve for Pd deviates strongly from the $1/N_{\rm PW}$ behavior except for very large values of $E_{\rm max}^{\chi}$. This is due to the properties of the highly localized $4d$ - electrons of Pd. Thus, for our purposes C can be considered to be an ``easy'' material, and Pd a ``difficult'' one. Heuristically, this can also be related to the fact that for the same range-separation parameter the relative amount of short-range correlation $E^{\rm RPA, SR}_{\rm c}/(E_{\rm c}^{\rm RPA, LR} + E_{\rm c}^{\rm RPA, SR})$ is smaller for C than for Pd. Likewise, total correlation energies are better reproduced for C than for Pd.

Due to systematic error cancellation, relative energies are in general easier to converge than total energies. To obtain equilibrium lattice constants, we have to compare systems of slightly different lattice volumes. We can expect good cancellation, if the orbitals of these systems are similar. A more detailed analysis will be given in the following.

\begin{table}[!tb] 
\begin{ruledtabular}
\caption{Cutoff energies $(Q_{\rm cut}+\Delta Q)^2/2$ for the cosine window ($\Delta Q = 0.1 Q_{\rm cut}$) and the SCK ($\Delta Q = 0.2 Q_{\rm cut}$). The range-separation parameters are given in $a_{\rm B}^{-1}$, the associated cutoff energies  are converted to eV.}
\label{tab:cutoffs}
\begin{tabular}{ccc}
$Q_{\rm cut}$ & $(Q_{\rm cut}+0.1Q_{\rm cut})^2/2$ & $(Q_{\rm cut}+0.2Q_{\rm cut})^2/2$\\ 
\hline \\\\[-3.\medskipamount]
2 & 66 & 78\\
3 & 148 & 176 \\
4 & 263 & 313
\end{tabular}
\end{ruledtabular}
\end{table}

\subsection{Lattice constants}

\begin{table}[!tb] 
\begin{ruledtabular}
\caption{Equilibrium lattice constants $a_0$ are given in \AA, errors (in \%) are given with respect to the zero point corrected experimental values, see Table \ref{tab:settings}. LDA and PBE calculations are performed self-consistently, full-range RPA and exact exchange are evaluated on top of self-consistent PBE orbitals.}
\label{tab:DFT_lattice_constants}
\begin{tabular}{lcccccc}
 & \multicolumn{2}{c}{LDA} & \multicolumn{2}{c}{PBE} & \multicolumn{2}{c}{RPA+EXX} \\
  \cline{2-3} \cline{4-5} \cline{6-7} \\\\[-3.\medskipamount]
& $a_0$ & Error & $a_0$ & Error & $a_0$ & Error \\
\hline \\\\[-3.\medskipamount]
C & 3.535 & -0.5 &3.571 & 0.5 & 3.565 & 0.3 \\ 
Si & 5.410 & -0.2 & 5.472 & 0.9 & 5.437 & 0.3 \\
MgO & 4.165 & -0.6 & 4.256 & 1.6 & 4.210 & 0.5\\
Kr   & 5.336  & -4.7 & 6.302 & 12.6 & 5.690 & 1.6 \\
Pd   & 3.842 & -0.9 & 3.940 & 1.7 & 3.895 & 0.5 \\
Al  & 3.982 & -0.9 & 4.035 & 0.4 & 4.036 & 0.4  
\end{tabular}
\end{ruledtabular}
\end{table}

\begin{figure}[!!tb]
\centering
\subfloat{\includegraphics [width=\linewidth,keepaspectratio=true] {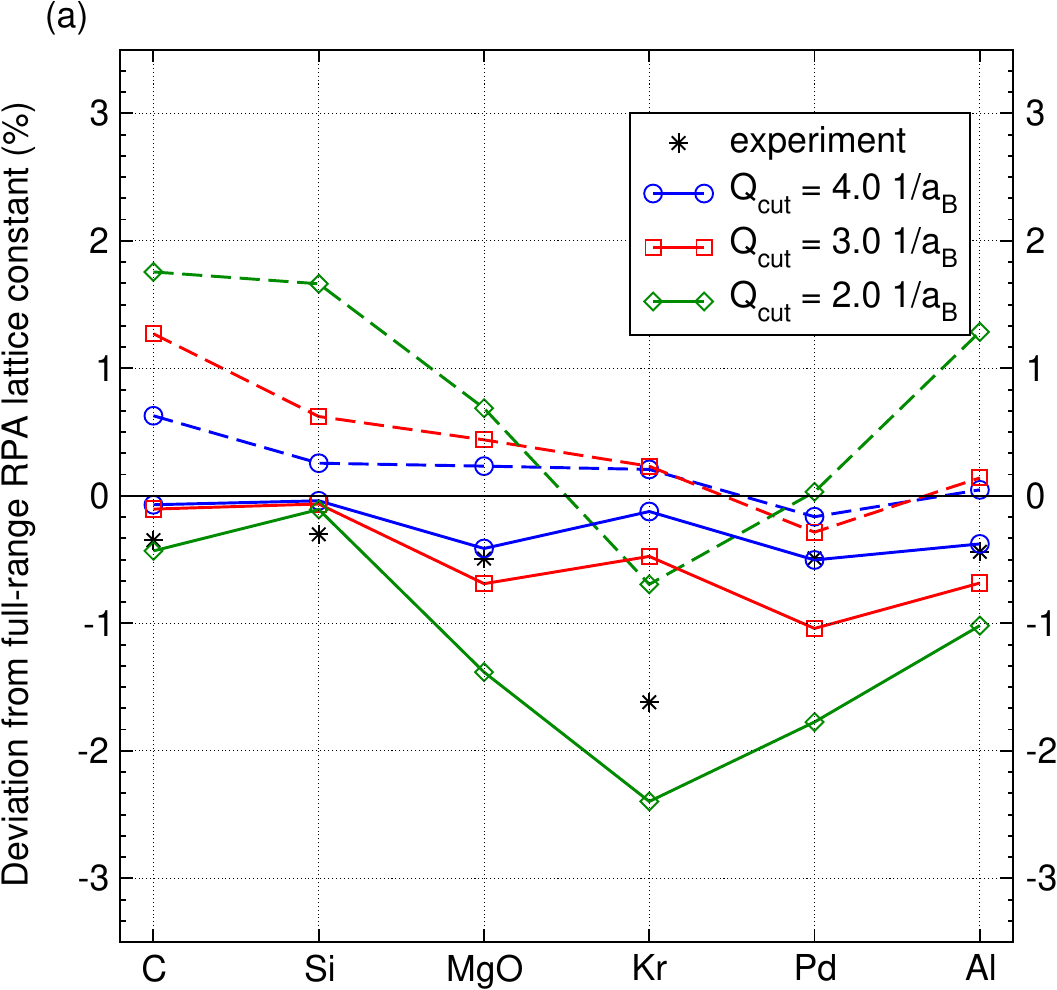}\label{fig:A6_LDA_a}}
\par\medskip
\subfloat{\includegraphics [angle=-90,width=\linewidth,keepaspectratio=true] {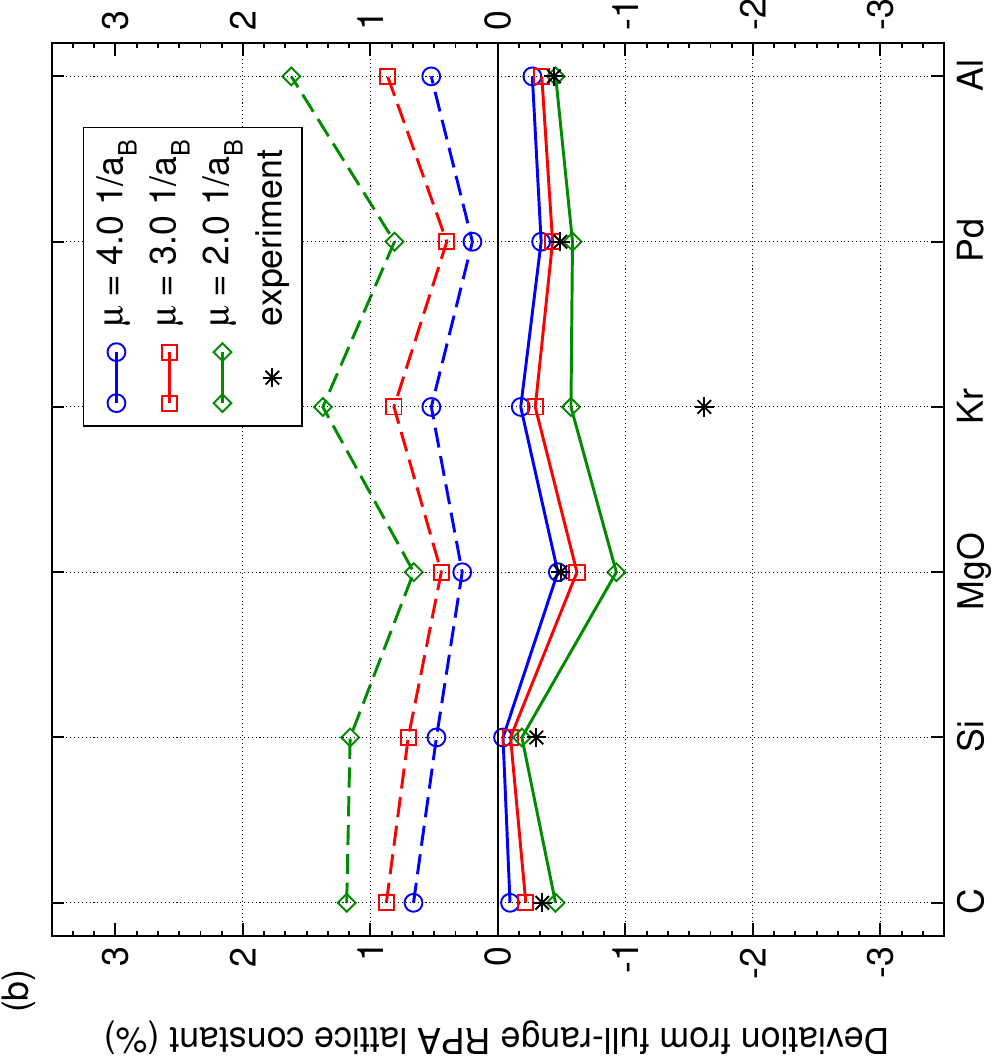}\label{fig:A6_LDA_b}}
\caption{Equilibrium lattice constants obtained via range-separated RPA correlation energies for \protect\subref{fig:A6_LDA_a} cosine window, \protect\subref{fig:A6_LDA_b} error function. Dashed lines represent results without, and solid lines represent results with short-range LDA correction. The full-range reference values are obtained with the extrapolation scheme \eqref{eq:Harl_extrapolation}.}
\label{fig:A6_LDA}
\end{figure}
 
\citet{Harl2010} have studied lattice constants for solids of various bonding types and found the following general behavior: LDA overbinds in comparison to the experimental values, PBE underbinds, and the RPA (with standard PAW potentials) underbinds, but less than PBE. This trend also applies to noble gas solids, though the (semi-)local functionals fail quite dramatically at describing van der Waals bonds with PBE yielding lattice constants that are significantly too large. \cite{Harl2008} Table \ref{tab:DFT_lattice_constants} confirms these results for the materials considered here. Slight differences to the full-range RPA lattice constants reported in the recent study of \citet{Klimes2015} (compare values listed as ``RPA std-PAW'') are due to higher cutoff energies used in the present work.

Fig. \ref{fig:A6_LDA} shows that replacing $E^{\rm RPA}_{\rm c} \rightarrow E_{\rm c}^{\rm RPA, LR}$ typically leads to larger lattice constants (dashed lines). The lattice constants tend to be overcorrected when $E^{\rm RPA, SR}_{\rm c}$ is included (solid lines).  In this section, the quality of the short-range LDA correction is only judged by {\em whether or not it yields lattice constants that are closer to the full-range RPA base line} than their respective long-range only counterparts (dashed lines), whereas in Sec. \ref{sec:semi-empirical RPA-LDA hybrid functional}, we will compare with experimental results. 

The performance of the short-range LDA correction is closely linked to the performance of the standard LDA, compare Table \ref{tab:DFT_lattice_constants}. The short-range correction works well for C and Si, where the difference between standard LDA and full-range RPA lattice constants is also smallest. Between those two materials, convergence with respect to the range-separation parameter is faster for Si, which is attributable to a size effect, i.e. $2k_{\rm F}$ is smaller for Si than for C.
For the other materials, however, the short-range LDA correction strongly overcorrelates. The most extreme cases are MgO, Pd and Kr in combination with the cosine window. At  $Q_{\rm cut} = 2 \: a_{\rm B}^{-1}$, i.e. $(Q_{\rm cut}+\Delta Q)^2/2 \approx 66$ eV, most of the relevant contributions to the correlation energy are simply disregarded in the RPA. In these cases, the LDA correction results in a quite severe underestimation of the lattice constant. The upshot is that an adequate cutoff must be at least around
$Q_{\rm cut} = 3 \: a_{\rm B}^{-1}$. 

Turning briefly to range-separation using the error function, we note that it is generally better behaved than the hard cosine cutoff,
but shares similar issues as the cosine window. The uncorrected lattice constants are too large, whereas the corrected lattice constants are a little bit too small. But ``stark'' 
outliners as for the cosine window are missing. We relate this to the fact that the error function cuts off the Coulomb kernel very slowly so that for momentum transfers $q\approx\mu$ 
a sizeable fraction of the Coulomb kernel is still present. This also means that one must use fairly large plane wave cutoffs for the response function to recover truly converged results for the correlation energy. Hence, computational gains are small when the error function is used for range-separation.

\begin{figure}[!!tb]
\centering
\includegraphics [angle=-90,width=\linewidth,keepaspectratio=true] {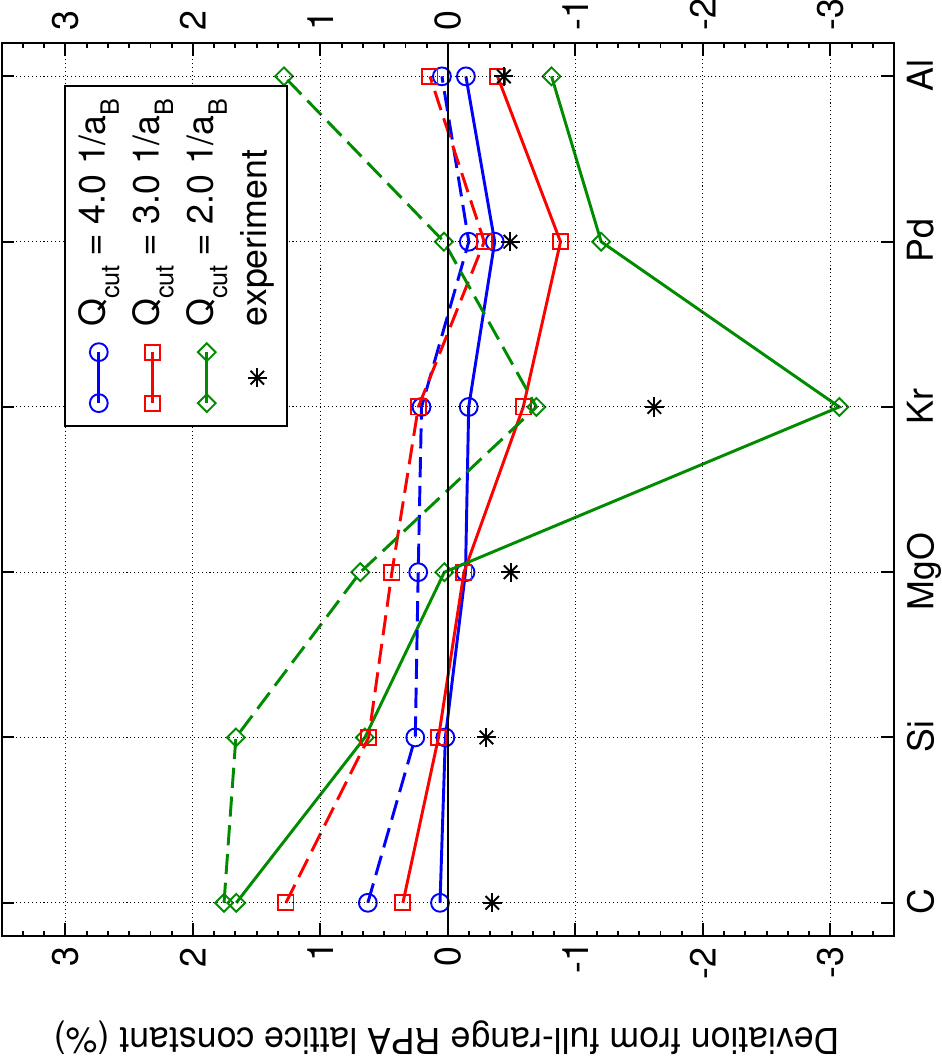}
\caption{Equilibrium lattice constants obtained via the SCK extrapolation method. Solid lines represent extrapolated results, dashed lines non-extrapolated results [same as in figure \ref{fig:A6_LDA}\protect\subref{fig:A6_LDA_a}]. The full-range reference values are obtained with the extrapolation scheme \eqref{eq:Harl_extrapolation}.}
\label{fig:A6_att}
\end{figure}

To judge the plane wave basis set extrapolation via the SCK, we compare to the non-extrapolated plane wave cutoff scheme, i.e. cosine windows using the same effective cutoff momentum $Q_{\rm cut}$. Fig. \ref{fig:A6_att} shows that the SCK shares the good description for C and Si with the short-range LDA correction, as well as the poor description for Kr and  Pd. This can be attributed to the success or failure of the HEG model, which is underlying both methods: if the correlation part that is not handled explicitly corresponds to plane wave-like contributions, both methods are reliable, otherwise the errors are sizable.  Hence, as for the cosine window,
results for $Q_{\rm cut} = 2 \: a_{\rm B}^{-1}$ are fairly unreliable. 

As discussed earlier, we expect that the SCK mimics some aspects of a short-range effective gradient correction, if the long-range effects are sufficiently well described. In fact, the SCK underbinds C and Si, but overbinds the other materials.  Noticeably, the MgO lattice constant is almost converged at $Q_{\rm cut} = 3 \: a_{\rm B}^{-1}$, and long-range correlation effects play an important role for MgO. However, the good description of MgO is maybe somewhat coincidental, as it again relies on fortuitous error cancellation, whereas the good description of C and Si also extends to total correlation energies (not shown here).  

\subsection{Semi-empirical RPA-LDA hybrid functional}\label{sec:semi-empirical RPA-LDA hybrid functional}

With decreasing range-separation parameter, the results for the error function change more slowly than those for  the cosine window. This is because the error function  mixes long- and short-range effects and does not abruptly cut off the Coulomb kernel. As discussed in Sec. \ref{sec:Riemelmoser2020_2}, we expect similar results for both methods for $Q_{\rm cut}= 2\mu$ in the DFT limit [compare the discussion following Eq. \eqref{eq:LR_exchange}]. In contrast, we identified that both methods work similar for $Q_{\rm cut}=\mu$ near the full-range limit, hence the slower change for the error function. The slow change of the results with respect to $\mu$  for the error function is in fact advantageous  for the construction of semi-empirical RPA-LDA hybrid functionals, where the optimal $\mu$ for one system should be transferable to other systems. 

For every material considered here, there is an optimal range-separation parameter $\mu$ or $Q_{\rm cut}$ that reproduces the experimental lattice constant. This is to be expected, since (i) the range-separated functional  switches between the full-range RPA and LDA. (ii) the overbinding of standard LDA is generally due to the correlation part, whereas the exchange part underbinds. (iii) LDA functionals based on the RPA instead of exact (QMC) data overbind even more than standard LDA, which can be deduced from the fact that the RPA+ correction typically increases the RPA lattice constants by 0.2-0.3\%. \cite{Harl2010}

Thus, by choosing the range-separation parameter empirically, we can improve the lattice constants upon full-range RPA. From our limited data set, we observe that the error function with $\mu \simeq 2k_{\rm F} $ gives the best lattice constants with respect to experiment [compare Fig. \ref{fig:A6_LDA}\protect\subref{fig:A6_LDA_b}]. Kr seems off, though the experimental value used here is possibly about 0.5\% too small. \cite{Rosciszewski2000}
For the cosine window, however, such a simple approximation does not work, since the results vary too rapidly as the cutoff changes. 

In summary, we find that the error function is better suited for the construction of semi-empirical RPA-LDA hybrid functionals. Even though the mixing of long- and short-range effects is not ideal in terms of removing the cusp-related UV divergence, it leads to improved transferability with respect to the optimal range-separation parameter. 

\subsection{Range-separated exchange}

\begin{figure}[!!tb]
\centering
\includegraphics [angle=-90,width=\linewidth,keepaspectratio=true] {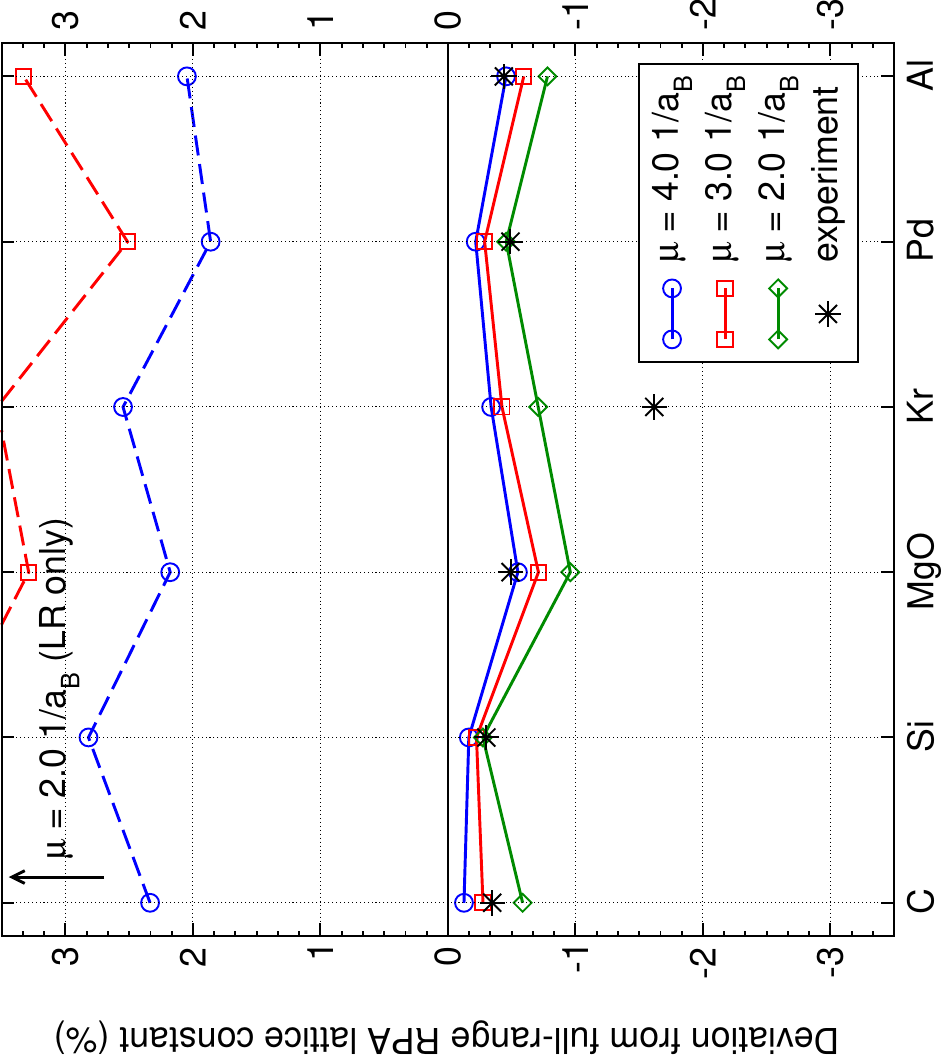}
\caption{Equilibrium lattice constants obtained via range-separated exchange and range-separated RPA correlation energies. Dashed line represent results without, and solid lines represent results with short-range LDA correction. The full-range reference values are obtained with the extrapolation scheme \eqref{eq:Harl_extrapolation}.}
\label{fig:A6_exchange}
\end{figure}

To study the influence of short-range exchange, we now replace the short-range part of the exchange energy by LDA similar to the treatment of correlation as described in section \ref{sec:Riemelmoser2020_2},
\begin{equation}
\begin{aligned} \label{eq:range_separated_exchange}
E_{\rm x} &= E_{\rm x}^{\rm LR} + E_{\rm x}^{\rm SR} \\
 &\approx E_{\rm x}^{\rm LR} + \int \text{d}\textbf{r} \; \varepsilon^{\rm SR}_{\rm x, HEG}[n(\textbf{r})]n(\textbf{r}) .
 \end{aligned}
\end{equation}
The treatment of $E_{\rm x}^{\rm LR}$ in VASP is technically elaborate, as one-center PAW terms are explicitly evaluated for exchange. Range-separated exchange has been implemented for the error function only and is rather involved for other Coulomb kernels. For details of the implementation in VASP we refer to Refs. \onlinecite{Paier2005} and \onlinecite{Angyan2006}. This problem does not occur for the RPA contribution, where the all-electron density is restored on the plane wave grid (see Ref. \onlinecite{Harl_thesis}). 

The error cancellation between exchange and correlation has played a central role in the success of standard LDA. This is manifest in the cancellation of the HEG terms linear in $r_{\rm s}$, as discussed earlier. The importance of joint treatment of exchange and correlation was also emphasized by \citet{Langreth1975} for their wave vector decomposition method. Joint treatment of range-separated exchange has also been employed by most of the authors investigating range-separated RPA based on the error function,\cite{Toulouse2009,Toulouse2010, Janesko2009,Janesko2009a,*Janesko2013,Janesko2009b,*Janesko2010} though none of these studies have commented on the role of short-range exchange on its own.

Fig. \ref{fig:A6_exchange} shows that entirely neglecting short-range exchange deteriorates the results even for large $\mu$. However, the short-range corrected lattice constants differ only slightly from the full-range exchange counterparts, compare Fig. \ref{fig:A6_LDA}. Thus, we can conclude that the LDA describes short-range exchange very well, in fact much better than short-range correlation.

In the following, we will comment on the observation that the short-range correction works better for exchange than for correlation near the full-range limit. This finding is not entirely surprising, since exchange is essentially a long-range effect. The exchange hole is quadratic in $r_{12}$ for small electronic distances $r_{12}$ (\emph{``no cusp for exchange''}). \cite{Toulouse2004} This is drastically manifest in the wave vector analysis of the HEG, where only momentum transfers $q \leq 2k_{\rm F}$ contribute [compare Eq. \eqref{eq:LR_exchange}]. For the error function, the separation between long- and short-range effects is not as extreme. As shown in Appendix \ref{App:Riemelmoser2020_A}, there is always at least a small short-range contribution [compare Eq. \eqref{eq:Savin_exchange}]. For large $\mu$ the short-range LDA exchange correction still becomes exact, as was shown originally by \myciteauthor{Gill1996},\cite{Gill1996}
\begin{equation}\label{eq:Gill}
E_{\rm x}^{\rm SR}(\mu) = -\frac{\pi}{4\mu^2}\int \text{d}\textbf{r} \; n(\textbf{r})^2 + \mathcal{O}\left(\frac{1}{\mu^4}\right)
\end{equation}
This is analogous to our theorem \eqref{eq:SR_LDA_theorem} on short-range RPA correlation, see also Appendix \ref{App:Riemelmoser2020_A}. But even beyond that, it seems plausible that the short-range LDA correction for exchange should work well near the full-range limit. There is no cusp for exchange, and so the short-range exchange contribution can be plane wave-like. In contrast, for correlation a large amount of plane waves is always needed to properly resolve the cusp.

It may still seem non-intuitive that the joint treatment of local exchange and local correlation is not beneficial here. However, we have focused on large $\mu$, i.e. on short-range effects, whereas the cancellation is valid for small $\mu$, compare Eqs. \eqref{NPexpansion} and \eqref{eq:LR_exchange}. Most clearly perhaps, this is seen in a real space picture. The cancellation of the HEG exchange and correlation holes pertains to the long-range parts, making the combined exchange-correlation hole more local. As for the short-range parts, we reiterate that there is a cusp for correlation but not for exchange.

Finally, the SCK method, which is designed to remove the electronic cusp explicitly, is not applicable to exchange. However, there exist similar methods for inverse range-separation, such as the cutoff scheme of \myciteauthor{Rozzi2006}\cite{Rozzi2006}

\section{\label{sec:Riemelmoser2020_5} Conclusion and Outlook}

We have shed new light on the plane wave basis set incompleteness error in the RPA  by relating it to  range-separated density functional theory. In a gist, the plane wave basis set incompleteness error of the RPA correlation energies can be  described by a complementary short-range local density functional.  In the limit of large cutoff energies, this partitioning becomes even exact. Furthermore, we have introduced a one-shot basis set correction method based on an optimized long-range potential. The method modifies the Coulomb kernel at intermediate wave vectors such that the truncated and ``squeezed'' Coulomb kernel reproduces the RPA correlation energy of the low-density homogeneous electron gas exactly. 

In practice, we find that the success of the two plane wave basis set correction methods used here is tied to that of the underlying HEG model. If the states that are 
removed from an explicit correlation treatment are plane wave-like, then the correction methods are successful. If the removed states are not plane wave-like, the correction
methods tend to fail or are at least less accurate. Difficult materials are obviously transition metals (here Pd), oxides (here MgO), but also in van der Waals
bonded solids (Kr) it is not a simple matter to develop accurate methods for correcting basis set errors. 

An interesting aspect that emerged from the present study is that since the RPA tends to underbind (underestimates relative binding energies, overestimates lattice
constants)  and the local density approximation overbinds (overestimates relative binding energies, underestimates lattice constants), there is always
one specific range-separation parameter that reproduces the experiment. Although the few data we have inspected suggest that $\mu \simeq 2k_{\rm F} $
yields the best results, also a fixed $\mu \simeq 3 \: a_{\rm B}^{-1}$ seems to work remarkably well for lattice constants. More extensive tests
will be required in order to tell whether such an approach will work equally well for other properties.

Somewhat surprisingly, we have found that the joint treatment of local exchange and local correlation does not improve the predicted lattice constants. As the short-range exchange correction is essentially exact for large $\mu$, the errors introduced in the correlation cannot be compensated by errors in the exchange. This can be interpreted by the fact that there is no cusp for exchange.

Finally, we note that the extrapolation via the squeezed Coulomb kernel can be further improved by including higher order terms in the low density expansion. Moreover, it would be desirable to generalize the method towards non-locality in order to describe more strongly correlated systems better. Climbing this Jacob's ladder for the optimized long-range potential should lead to faster convergence with respect to the plane wave basis set size. 

\section*{Acknowledgements}

The authors wish to express their gratitude to F. Hummel for support. Computation time at the Vienna Scientific Cluster (VSC) is gratefully acknowledged. 

\begin{appendix}
\section{Range-separated exchange}\label{App:Riemelmoser2020_A}

The exchange energy per particle of an HEG interacting via a long-range interaction $V^{\rm LR}$ is given as
\begin{equation}
\begin{aligned}
\varepsilon_{\rm x, HEG}^{\rm LR} (r_{\rm s}) = 
-\Omega \int \frac{\text{d}\textbf{q}}{(2\pi)^3} V^{\rm LR} (|\textbf{q}|) \\\times \int \frac{\text{d}\textbf{k}}{(2\pi)^3} \theta(k_{\rm F}-|\textbf{k}+\textbf{q}|)\theta(k_{\rm F}-|\textbf{k}|)
\end{aligned}
\end{equation}
The integration over $\textbf{k}$ yields (compare derivation of \myciteauthor{Fetter2003},\cite{Fetter2003} chap. 3)
\begin{equation}
\begin{aligned}
&\varepsilon_{\rm x, HEG}^{\rm LR} (r_{\rm s}) = \\& -\frac{2k_{\rm F}^3}{\pi^2}  \int_0^1 \text{d}y \; y^2 V^{\rm LR}(y) \left(1 - \frac{3}{2}y+\frac{1}{2}y^3\right) , 
\end{aligned}
\end{equation}
where $y = q/2k_{\rm F}$. 

For the hard cutoff \eqref{eq:hardcutoff}, we obtain Eq. \eqref{eq:LR_exchange}, and for the error function \eqref{eq:error_function} we recover the result of \citet{Savin1996}
\begin{equation} \label{eq:Savin_exchange}
\begin{aligned}
\varepsilon_{\rm x, HEG}^{\rm LR}(\mu,r_{\rm s}) =& -\frac{\mu}{\pi}\bigg[(2x-4x^3)e^{-1/4x^2}\\
&-3x+4x^3+\sqrt{\pi}\text{erf} \left(\frac{1}{2x}\right)\bigg] ,
\end{aligned}
\end{equation}
where $x=\mu/2k_{\rm F}$. For large $\mu$, this can be expanded as
\begin{equation}
\varepsilon_{\rm x, HEG}^{\rm LR}(\mu,r_{\rm s}) = \varepsilon_{\rm x} + \frac{3}{16r_{\rm s}^3\mu^2} + \mathcal{O}\left(\frac{1}{\mu^4}\right) .
\end{equation}
By inserting this into equation \eqref{eq:range_separated_exchange}, we obtain an expression for the LDA approximation to the short-range exchange energy
\begin{equation}
E_{\rm x}^{\rm SR, LDA}(\mu) = -\frac{\pi}{4\mu^2}\int \text{d}\textbf{r} \; n(\textbf{r})^2 + \mathcal{O}\left(\frac{1}{\mu^4}\right) .
\end{equation}
\citet{Gill1996} have shown that, up to leading order, this is also the expression for the exact short-range exchange, compare Eq. \eqref{eq:Gill}.	

\section{Short-range RPA energy for the HEG: \\
Approaching the full-range limit}\label{App:Riemelmoser2020_B}

Our derivation of the high $Q_{\rm cut}$-limit of $\varepsilon_{\rm c, HEG}^{\rm RPA, SR}(Q_{\rm cut})$ follows the study of \myciteauthor{Gulans2014}.\cite{Gulans2014} The text-book equation for full-range expression reads \cite{Ren2012}
%
\begin{equation}
\begin{aligned}
&\varepsilon^{\rm RPA}_{\rm c, HEG}
= \frac{1}{n} \int \frac{\text{d}q}{(2\pi)^3} 4\pi q^2 \int_{0}^{\infty}\frac{\text{d}\omega}{2\pi}\\
&\times \ln [\left(1-\chi_{0, \rm HEG}(q,i\omega)V(q) \right)
 + \chi_{0, \rm HEG}(q,i\omega)V(q)] ,
 \end{aligned}
\end{equation}
%
where $\chi_{0, \rm HEG}(q,i\omega)$ is the Lindhard polarizibility for the HEG in terms of imaginary frequencies. The long-range version is obtained by replacing $V \rightarrow V^{\rm LR}(Q_{\rm cut})$ [see Eq. \eqref{eq:hardcutoff}], and $\varepsilon_{ \rm c, HEG}^{\rm RPA, SR}(Q_{\rm cut})$ is given by the difference between those. It is convenient to express $\chi_{0, \rm HEG}$ in the closed form (see Ref. \onlinecite[chap. 4]{Giuliani2008})
\begin{equation}\label{eq:Giuliani_Lindhard}
\begin{aligned}
\chi_{0, \rm HEG}(q,i\omega) =  \frac{k_{\rm F}^2}{\pi^2q}\bigg[\Psi\left( \frac{i\omega}{qk_{\rm F}}-\frac{q}{2k_{\rm F}} \right)\\
-\Psi\left( \frac{i\omega}{qk_{\rm F}}+\frac{q}{2k_{\rm F}}\right)\bigg] ,
\end{aligned}
\end{equation}
where $\Psi(z)$ is defined as 
\begin{equation}\label{eq:Giuliani}
\Psi(z) = \frac{z}{2} + \frac{1-z^2}{4}\ln\left(\frac{z+1}{z-1}\right).
\end{equation}
Note that Gulans missed a factor of 2 in his expression for $\chi_{0,\rm HEG}$, that accounts for summation over spin. This also affects his result for the $GW$ basis set incompleteness error [compare Eq. (28) in Ref. \onlinecite{Klimes2014a}]. We introduce the dimensionless variables $y=q/2k_{\rm F}$ and $t=2\omega/q^2$ and use the series expansion
\begin{equation}
\Psi(z) \overset{|z| \to \infty}{=} \frac{1}{3z} + \frac{1}{15z^3} + \frac{1}{35z^5} + \mathcal{O}\left(\frac{1}{z^7}\right).
\end{equation}
Hence, for large momentum transfers $q$ we obtain
\begin{equation}
\begin{aligned}
&\chi_{0, \rm HEG} (y,t) =\\
& -\frac{k_{\rm F}}{2\pi^2y^2}\bigg[\frac{2}{3}\frac{1}{1+t^2} + \frac{2}{15y^2}\frac{1-3t^2}{(1+t^2)^3} \\
&+ \frac{2}{35y^4} \frac{1-10t^2+5t^4}{(1+t^2)^5} +  \mathcal{O}\left(\frac{1}{y^6}\right)\bigg] .
\end{aligned}
\end{equation}
In this limit, we may also use the Mercator expansion
\begin{equation}
\ln(1-x) = -x - \frac{x^2}{2}-\frac{x^3}{3}-... \hspace{15pt} \text{for } x \to 0,
\end{equation}
to decompose the RPA into its ring diagram components, compare Fig. \ref{fig:diagrams}. We proceed to evaluate the frequency integrations, starting with the direct MP2 part
\begin{equation}
\begin{aligned}
&\int_0^\infty \frac{\text{d}\omega}{2\pi} \frac{\left(\chi_{0,\rm HEG}V)^2 -(\chi_{0, \rm HEG} V^{\rm LR}\right)^2}{2} = \\&\frac{\theta(q-Q_{\rm cut})}{8\pi^2y^6} \left[\frac{1}{9}+\frac{1}{90y^2}+\frac{1}{350y^4} + \mathcal{O}\left(\frac{1}{y^6}\right)\right] .
\end{aligned}
\end{equation}
After momentum integration, we find that the second order contribution to $\varepsilon_{\rm c, HEG}^{\rm RPA,SR}$ is given by 
\begin{equation}
\begin{aligned}
&\varepsilon_{\rm c, HEG}^{\rm RPA,SR,(2)}(Q_{\rm cut}) =\\ 
&-\frac{1}{\pi}\bigg[\frac{1}{Q_{\rm cut}^3r_{\rm s}^3} + \frac{6}{25\alpha^2Q_{\rm cut}^5r_{\rm s}^5}\\
& + \frac{216}{1225\alpha^4Q_{\rm cut}^7r_{\rm s}^7} + \mathcal{O}\left(\frac{1}{Q_{\rm cut}^9}\right) \bigg] .
\end{aligned}
\end{equation}
For contributions from the third order ring diagram, we evaluate

\begin{equation}
\begin{aligned}
&\int_0^\infty \frac{\text{d}\omega}{2\pi} \frac{\left(\chi_{0, \rm HEG}V)^3-(\chi_{0, \rm HEG} V^{\rm LR}\right)^3}{3} =\\ &- \frac{\theta(q-Q_{\rm cut})}{432\pi^3k_{\rm F}y^{10}} + \mathcal{O}\left(\frac{1}{y^{12}}\right), 
\end{aligned}
\end{equation}
and obtain for the short-range correlation energy per particle

\begin{equation}
\varepsilon_{\rm c, HEG}^{\rm RPA,SR,(3)}(Q_{\rm cut}) = \frac{18}{7\pi Q_{\rm cut}^7r_{\rm s}^6} + \mathcal{O}\left(\frac{1}{Q_{\rm cut}^9}\right) .
\end{equation}
Contributions of higher order diagrams are $\mathcal{O}(1/Q_{\rm cut}^9)$ as well. The last equation shows that the low-density scaling behavior starts to break down at $\mathcal{O}(1/Q_{\rm cut}^7)$ through the influence of higher order diagrams. 

\section{Fitting the short-range LDA functionals}\label{App:Riemelmoser2020_C}

\begin{figure}[htb]
\centering
\includegraphics [width=\linewidth,keepaspectratio=true] {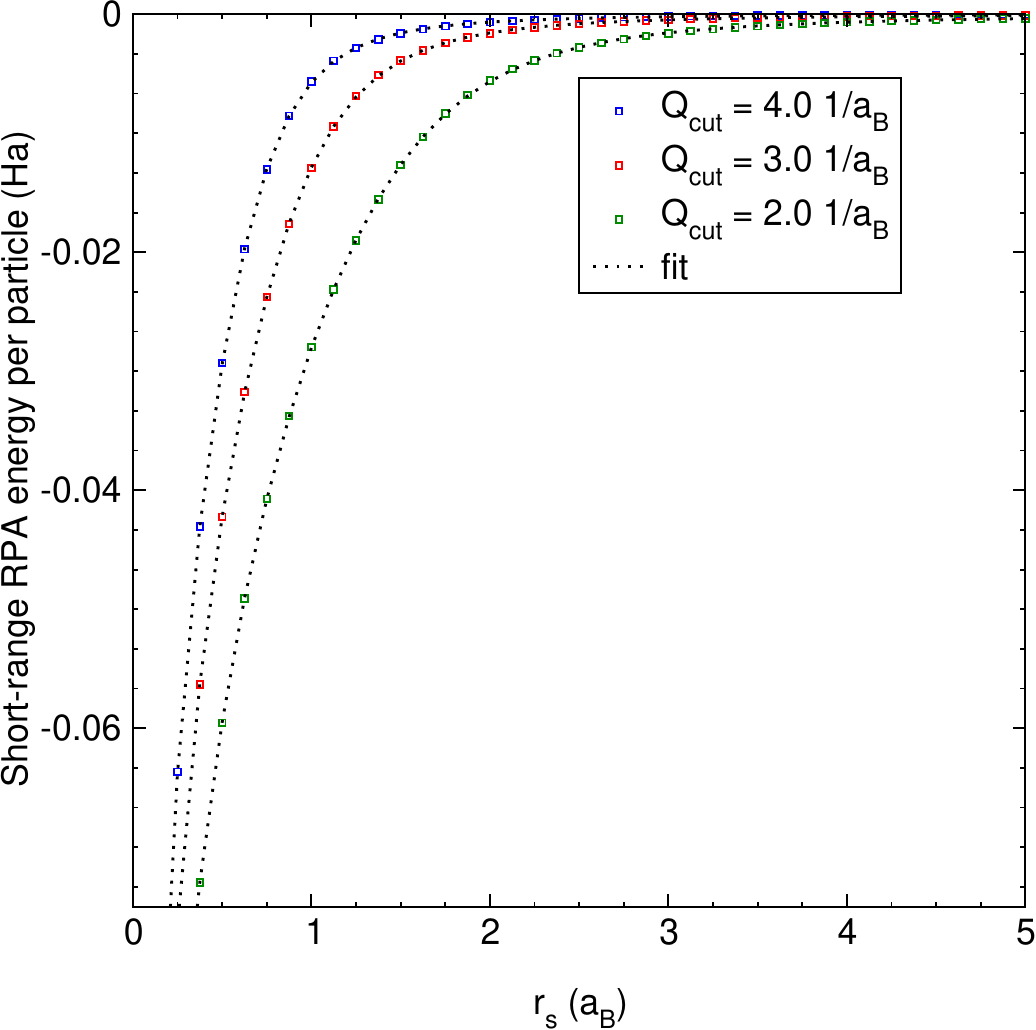}
\caption{Analytic representation of the short-range LDA functionals for the cosine window. Square symbols represent numerical data for different range-separation parameters, and dotted lines represent fits using Eq. \eqref{eq:pade_fit_parameters}.}
\label{fig:fit_cos}
\end{figure}

\begin{figure}[htb]
\centering
\includegraphics [angle=-90,width=\linewidth,keepaspectratio=true] {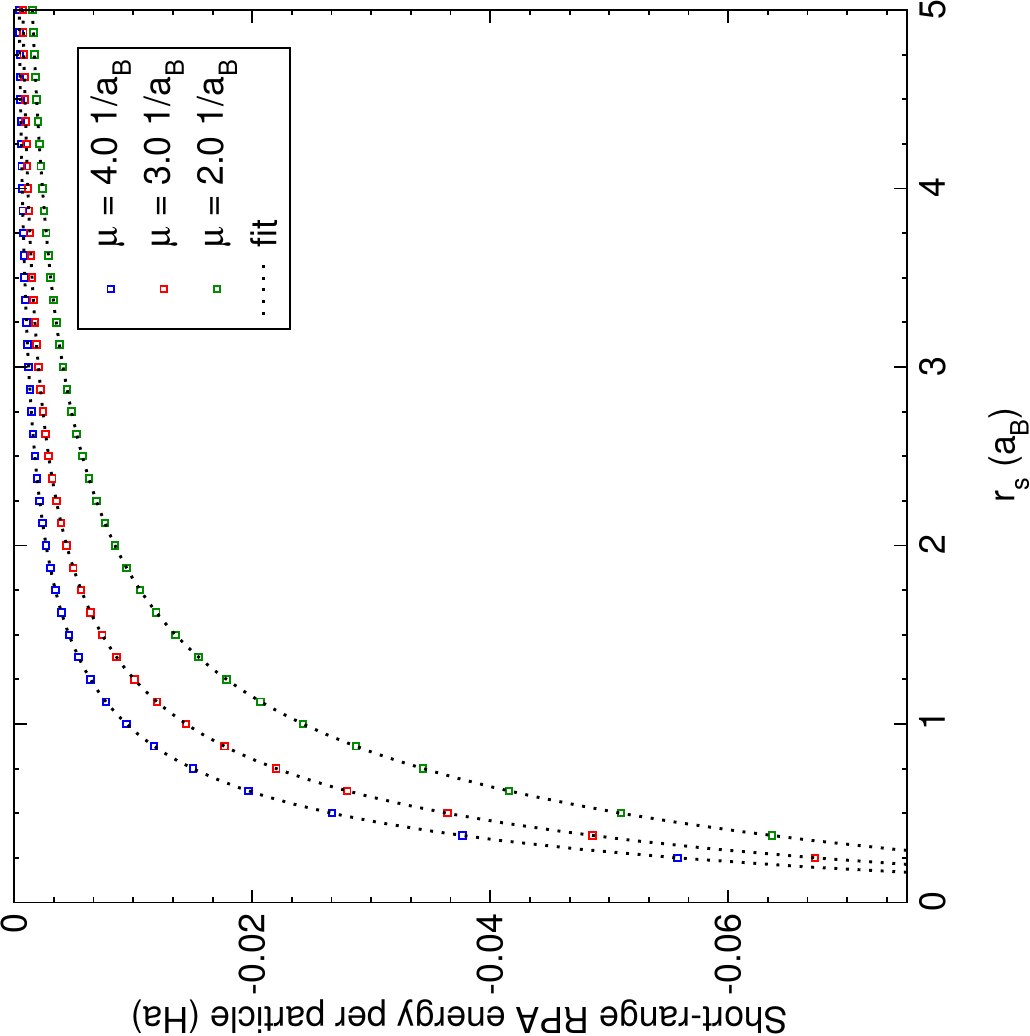}
\caption{Analytic representation of the short-range LDA functionals for the error function. Square symbols represent numerical data for different range-separation parameters, and dotted lines represent fits using Eq. \eqref{eq:pade_fit_parameters}.}
\label{fig:fit_mu}
\end{figure}

As outlined in section \ref{sec:Riemelmoser2020_3}, we fit the short-range LDA functionals using the form
\begin{equation}\label{eq:pade_fit_parameters}
\begin{aligned}
&\varepsilon_{\rm c}^{\rm RPA, SR} = A \frac{\ln \left(\frac{r_{\rm s} + a_0 r_{\rm s}^2 + a_1 r_{\rm s}^3 + a_2 r_{\rm s}^4}{1+a_3 r_{\rm s} + a_4 r_{\rm s}^2 + a_5 r_{\rm s}^3 + a_2 r_{\rm s}^4}\right)}{1+a_6 r_{\rm s} + a_7 r_{\rm s}^2} \\
&A = - \frac{\ln(2)-1}{\pi^2} .
\end{aligned}
\end{equation}
In Tables \ref{tab:parameters_cos} and \ref{tab:parameters_mu}, the fit parameters are given for selected range-separation parameters.
Figs. \ref{fig:fit_cos} and \ref{fig:fit_mu} demonstrate the fits for the cosine window and the error function respectively.

\clearpage

\begin{table}[htb] 
\begin{ruledtabular}
\caption{Fit parameters for the cosine window. The Wigner-Seitz radius $r_{\rm s}$ is input in $a_{\rm B}$, the short-range RPA energy per particle $\varepsilon_{\rm c}^{\rm RPA, SR}$ is output in Ha.}
\label{tab:parameters_cos}
\begin{tabular}{crrr}
& $Q_{\rm cut}= 2 \; a_{\rm B}^{-1}$ & $Q_{\rm cut}= 3 \; a_{\rm B}^{-1}$ & $Q_{\rm cut}= 4 \; a_{\rm B}^{-1}$  \\
\hline \\\\[-3.\medskipamount]
$a_0$ &723.273 & 250.439  & 42.2121 \\
$a_1$ & -778.762 & -458.185 & -115.400 \\
$a_2$ & 434.396 & 368.688  & 117.648 \\
$a_3$ & 6985.71 & 2192.95 & 371.181 \\
$a_4$ & -2873.23 & -1452.77 & -347.389 \\
$a_5$ & 251.151  & 295.871 & 124.814 \\
$a_6$ & 0.958156 & 1.53924 & 1.88767 \\
$a_7$ & 0.854852 & 2.67992 & 6.72314 \\
\end{tabular}
\end{ruledtabular}
\end{table}

\begin{table}[htb] 
\begin{ruledtabular}
\caption{Fit parameters for the error function. The Wigner-Seitz radius $r_{\rm s}$ is input in $a_{\rm B}$, the short-range RPA energy per particle $\varepsilon_{\rm c}^{\rm RPA, SR}$ is output in Ha.}
\label{tab:parameters_mu}
\begin{tabular}{crrr}
& $\mu = 2 \; a_{\rm B}^{-1}$ & $\mu = 3 \; a_{\rm B}^{-1}$ & $\mu = 4 \; a_{\rm B}^{-1}$  \\
\hline \\\\[-3.\medskipamount]
$a_0$ & 60.2614 & 26.6952 & 56.7518 \\
$a_1$ & -50.7152 & -38.9317 & -113.350 \\
$a_2$ & 141.086 & 138.271 & 523.105 \\
$a_3$ & 728.749 & 439.932 & 703.130 \\
$a_4$ & 722.861 & 458.791 & 1003.12 \\
$a_5$ & 409.769 & 351.941 & 1002.40 \\
$a_6$ & 2.26403 & 4.04404 & 5.21364 \\
$a_7$ & 0.0416747 & 0.104055 & 0.157932 \\
\end{tabular}
\end{ruledtabular}
\end{table}

\end{appendix}
\section*{References}

\bibliography{Riemelmoser2020_postprint}
\printindex

\end{document}